\documentclass[11pt]{article}

\usepackage[final]{acl}

\usepackage{times}
\usepackage{latexsym}
\usepackage[T1]{fontenc}
\usepackage[utf8]{inputenc}
\usepackage{microtype}
\usepackage{inconsolata}
\usepackage{graphicx}
\usepackage{booktabs}
\usepackage{multirow}
\usepackage{amsfonts}
\usepackage{amsmath}
\usepackage{tabularx}
\usepackage{array}

\title{\textsc{Tempo}: \underline{Te}mporally-grounded \underline{M}ulti-task \underline{Po}st-training for Large Audio-Language Models}

\renewcommand{\thefootnote}{\fnsymbol{footnote}}

\author{
  \textbf{Apoorva Kulkarni\textsuperscript{1}}\footnotemark[1],
  \textbf{Kaousheik Jayakumar\textsuperscript{1}}\footnotemark[1],
  \textbf{Sreyan Ghosh\textsuperscript{1}},
  \textbf{Utathya Aich\textsuperscript{2}}, \\
  \textbf{Ramani Duraiswami\textsuperscript{1}}\footnotemark[2],
  \textbf{Dinesh Manocha\textsuperscript{1}}\footnotemark[2]
\\
\\
  \textsuperscript{1}University of Maryland, College Park, USA \quad \textsuperscript{2}CNH Industrial India
\\
  \textbf{Correspondence:} \href{mailto:apoorvak@umd.edu}{apoorvak@umd.edu} \\
  \centering Project: \url{https://kaousheik-26.github.io/tempo/}
}

\begin{document}
\maketitle

\begingroup
\renewcommand{\thefootnote}{}
\footnotetext{$^{*}$Equal Contribution. \quad $^{\dagger}$Equal Advising.}
\endgroup

\renewcommand{\thefootnote}{\arabic{footnote}}
\setcounter{footnote}{0}

\begin{abstract}
Large audio-language models (LALMs) describe audio at the clip level but cannot assign timestamps to the events, speakers, or sounds they identify. Despite being essential for downstream tasks like speech recognition and dense audio captioning, timestamping remains a key limitation of most LALMs. We present \textsc{Tempo} (\textbf{Te}mporally-grounded \textbf{M}ulti-task \textbf{Po}st-training), the first unified model 
to handle audio, speech and music timestamping tasks. Our core contribution is a supervised fine-tuning (SFT) stage built on three innovations: atomic timestamp tokens, a time-aware projector that injects sinusoidal wall-clock encodings into audio frame embeddings, and a distance-aware Gaussian loss. Our training is based on a synthetic-to-real curriculum. We further introduce, to our knowledge, the first application of
reinforcement learning to unified audio timestamping, using GRPO with verifiable temporal rewards that directly optimize the evaluation objectives. Rather than serving as the primary source of performance gains, GRPO acts as a refinement stage on top of the SFT checkpoint, providing modest additional improvements. To support this work, we build a training dataset containing 119K samples and an evaluation benchmark containing 10K samples, drawn from established corpora across five tasks. On this benchmark, \textsc{Tempo} outperforms Audio Flamingo Next and Qwen3-Omni, two state-of-the-art LALMs explicitly trained on timestamped data. Experiments confirm that SFT delivers most of these gains, with GRPO providing consistent but moderate refinements. 
\end{abstract} 

\section{Introduction}

Audio signals carry rich temporal information. Answering \emph{what happens when} in an audio stream is fundamental to applications such as meeting transcription and music analysis, which require associating specific events with precise time intervals. Yet most existing Large Audio-Language Models (LALMs) treat audio understanding as a clip-level task, mapping a recording to a single caption or answer and discarding temporal structure~\citep{ghosh2024gamalargeaudiolanguagemodel,goel2025audioflamingo3advancing,kong2024audioflamingonovelaudio,ghosh2025audioflamingo2audiolanguage}. Timestamping is therefore a foundational capability that current LALMs lack.

In this work, we focus on five timestamping tasks that span speech, music and audio: multi-speaker automatic speech recognition (ASR), speaker diarization, audio temporal grounding, dense audio captioning and timestamped music captioning. Despite differing in their outputs, all five share a common structure: each requires segmenting an audio into intervals and assigning a textual label to each one. Recent LALM-based approaches such as TAC~\citep{kumar2026tactimestampedaudiocaptioning} and TimeAudio~\citep{wang2025listeningframesbridgingtemporal} address timestamping within a unified model but remain restricted to limited tasks.

We present \textsc{Tempo}, the first unified LALM for timestamping across audio, speech, and music. 
\textsc{Tempo} handles all five tasks with a single decoder, emitting text with timestamp tokens distinguished only by a task tag in the prompt. Our supervised fine-tuning (SFT) recipe builds on three components: atomic timestamp tokens, a time-aware multi-modal projector that injects sinusoidal wall-clock encodings into audio frame embeddings and a distance-aware Gaussian loss. Training follows a two-stage synthetic-to-real curriculum. Building on this, we present the first application of reinforcement learning to timestamped audio understanding, using Group Relative Policy Optimization (GRPO)~\citep{Guo_2025} with verifiable rewards to train directly against the temporal metrics used at evaluation. 

On a benchmark constructed from established corpora, \textsc{Tempo} substantially
outperforms Audio Flamingo Next and Qwen3-Omni, two state-of-the-art LALMs explicitly
trained on timestamped data. Gains are largest on speech: \textsc{Tempo} cuts WER from
0.70 to 0.44 on multi-speaker ASR and improves diarization mIoU from 0.44 to 0.71 over
Qwen3-Omni, with consistent improvements across audio and music tasks as well. Experiments confirm that SFT delivers most of these gains, with GRPO providing consistent but moderate refinements, suggesting that careful post-training design is the primary lever for adding timestamping capabilities to LALMs.
Our main contributions include:
\begin{itemize}
    \item \textbf{First unified LALM for timestamping across speech, sound, and music.} \textsc{Tempo} performs speaker diarization, multi-speaker ASR, audio temporal grounding, dense audio captioning, and timestamped music captioning in a single model, outperforming Qwen3-Omni, Audio Flamingo Next and other LALM-based timestamping baselines.
    \item \textbf{A post-training SFT recipe that teaches timestamping from scratch.} Three innovations, namely atomic timestamp tokens at 0.1\,s resolution, a time-aware multi-modal projector with sinusoidal wall-clock encodings, and a distance-aware Gaussian loss, together teach a model with no prior temporal supervision to predict accurate timestamps.
   \item \textbf{First unified RL approach across speech, audio, and music timestamping.} We apply GRPO with task-specific verifiable temporal
rewards jointly across speech, sound, and music. Starting from the SFT
checkpoint, GRPO serves as a refinement that directly optimizes
the evaluation objectives, yielding modest additional gains on the speech and sound tasks.
    \item \textbf{Multi-task benchmark and training corpus.} 119K training examples (51K synthetic, 68K real-world) and 10K evaluation examples drawn from established corpora, covering all five timestamping tasks under a unified output format.
\end{itemize}

\section{Related Work}

\subsection{Large Audio-Language Models and Timestamping}

Recent LALMs including SALMONN \citep{tang2024salmonngenerichearingabilities}, Qwen2-Audio \citep{chu2024qwen2audiotechnicalreport}, the Audio Flamingo series~\citep{kong2024audioflamingonovelaudio,ghosh2025audioflamingo2audiolanguage,goel2025audioflamingo3advancing,ghosh2026audioflamingonextnextgeneration}, GAMA~\citep{ghosh2024gamalargeaudiolanguagemodel}, and Kimi-Audio~\citep{kimiteam2025kimiaudiotechnicalreport} have demonstrated strong capabilities on captioning, question answering, and open-ended reasoning, but temporal grounding remains a persistent weakness. Recent studies further show that even strong LALMs struggle with fine-grained temporal reasoning, including event ordering, duration, and onset/offset estimation~\citep{bhattacharya2025benchmarkingconfidenceevaluationlalms}, \citep{kulkarni2026closerlookfailuremodes}. These findings further motivate explicit temporal supervision beyond clip-level audio understanding. The closest prior work, TAC~\citep{kumar2026tactimestampedaudiocaptioning} and TimeAudio~\citep{wang2025listeningframesbridgingtemporal}, address timestamping within a unified model but are restricted to limited tasks and rely on supervised training. \textsc{Tempo} differs from its closest prior work, TAC, in both
training data and task coverage. TAC uses only synthetic data for dense
audio captioning and delegates speech transcription to
Whisper~\citep{radford2023robust}, whereas \textsc{Tempo} uses publicly
available human-annotated data and jointly models five tasks spanning
speech, sound, and music. Consequently, TAC does not support
speaker-attributed multi-speaker ASR or speaker diarization. TAC also
identifies timestamped chord progression generation as a limitation,
whereas \textsc{Tempo} directly predicts structured timestamped chord
progressions. We do not report a direct numerical comparison because
public TAC model weights and inference code are unavailable.

\begin{table*}[t]
\centering
\resizebox{\textwidth}{!}{
\begin{tabular}{|l|p{4.5cm}|p{10cm}|}
\hline
\textbf{Task} & \textbf{Question} & \textbf{Ground Truth Response} \\
\hline
ASR &
Provide a timestamped transcription of the speech in this audio. &
\texttt{[speech:asr] <|0.0|> The memory uh <|2.2|>}
\texttt{[speech:asr] <|3.4|> Pops up the options <|4.5|>}
\texttt{[speech:asr] <|4.9|> Yeah yeah that would be possible yeah sure <|7.1|>}
\texttt{[speech:asr] <|7.5|> I th dont think thats uh that takes a lot of storage space \ldots <|14.0|>} \\
\hline
Diarization &
How many speakers are in this audio and when does each one speak? Include timestamps. &
\texttt{[speech:diar] <|0.0|> Speaker 1 <|6.2|>}
\texttt{[speech:diar] <|4.6|> Speaker 2 <|6.1|>}
\texttt{[speech:diar] <|5.9|> Speaker 3 <|16.9|>}
\texttt{[speech:diar] <|13.5|> Speaker 1 <|27.9|>} \\
\hline
Audio Grounding &
What is the time interval (start and end) for the query `a bird tweets quietly and repeatedly outdoors' in the audio? &
\texttt{[audio:ground] <|0.9|> to <|1.4|>}
\texttt{[audio:ground] <|2.3|> to <|2.8|>}
\texttt{[audio:ground] <|3.7|> to <|4.3|>}
\texttt{[audio:ground] <|5.2|> to <|5.7|>} \\
\hline
Dense Audio Captioning &
Which sound events occur, and what are their time intervals and descriptions? &
\texttt{[audio:caption] <|0.0|> A river flows loudly outdoors. <|23.9|>}
\texttt{[audio:caption] <|0.9|> A bird tweets quietly and repeatedly outdoors. <|1.4|>}
\texttt{[audio:caption] <|2.3|> A bird tweets quietly and repeatedly outdoors. <|2.8|>}
\texttt{[audio:caption] <|3.7|> A bird tweets quietly \ldots <|4.3|>} \\
\hline
Music Captioning &
Caption this music with timestamps. Describe the instrument, tempo, chord progression, dynamics, and note statistics. &
\texttt{[audio:music][instrument] Organ enters at <|0.0|> exits at <|30.0|>}
\texttt{[tempo] 120.0 BPM at <|0.0|>}
\texttt{[tempo] 125.0 BPM at <|9.5|>}
\texttt{[chord] B:min from <|0.0|> to <|3.5|>}
\texttt{[chord] D:sus2 from <|3.5|> to <|21.0|>}
\texttt{[chord] C:sus2 from <|21.0|> to <|28.5|>}
\texttt{[chord] D:min7 from <|28.5|> to <|29.0|>}
\texttt{[chord] G:sus2 from <|29.0|> to <|29.5|>}
\texttt{[chord] D:sus2 from <|29.5|> to <|30.0|>}
\texttt{[stats] Note density:0.6 notes/sec, range: D5 to G6} \\
\hline
\end{tabular}
}
\caption{Example instances from each evaluation task. Ground truth responses are truncated for brevity.}
\label{tab:data_examples}
\end{table*}
\subsection{Speech: Diarization and Multi-Speaker ASR}

Diarization has evolved from clustering pipelines to end-to-end and target-speaker models; multi-speaker ASR extends this with joint transcription via serialized output training. Recent LLM-based methods include DiarizationLM~\citep{diarizationLM}, which refines outputs via post-processing and SpeakerLM~\citep{yin2026speakerlmendtoendversatilespeaker} jointly models speaker and language. Unlike these speech-only systems, \textsc{Tempo} jointly predicts \emph{who}, \emph{what}, and \emph{when} within a general-purpose framework trained across non-speech tasks as well.

\subsection{Sound: Event Detection, Temporal Grounding and Dense Captioning}

Sound event detection (SED) and the DCASE challenge series provide established formulations and evaluation protocols for temporally localizing acoustic events~\citep{Mesaros_2021, Politis_2021}. Recent work has also connected SED with language models for generating event labels and temporal locations~\citep{wang2024leveraginglanguagemodelcapabilities}, \citep{yang2026multidomainaudioquestionanswering}. Audio grounding localizes a text query to a time interval, while dense captioning requires segmenting an audio stream and describing each event with timestamps. Standard benchmarks like AudioCaps~\citep{audiocaps} and Clotho~\citep{drossos2019clothoaudiocaptioningdataset} provide only clip-level annotations, inducing \emph{semantic collapse} where distinct events are compressed into global summaries~\citep{kumar2026tactimestampedaudiocaptioning}. Recent efforts address this via synthetic dense annotation pipelines and frame-level temporal indexing~\citep{wang2025listeningframesbridgingtemporal}.

\subsection{Music: Chord Estimation and Captioning}

Automatic chord estimation traditionally combines acoustic classifiers with harmonic
language models, evaluated on McGill Billboard~\citep{burgoyne2011expert} using
MIREX-standard metrics, but these systems are specialized classifiers that do not
generalize beyond chord recognition. Music Flamingo~\citep{ghosh2025musicflamingoscalingmusic}
introduces rotary time embeddings for temporal localization and produces theory-aware
captions that may reference chord progressions in natural language, but does not generate
structured, timestamped chord labels. To our knowledge, no existing audio-language model
produces structured, timestamped chord progressions from audio; we integrate this as one
of five tasks in \textsc{Tempo}.

\subsection{Reinforcement Learning for LALMs}

The DeepSeek-R1 pipeline~\citep{Guo_2025}, which combines an SFT cold start with Group Relative Policy Optimization (GRPO)~\citep{shao2024deepseekmath} and verifiable rewards, has proven effective for text and vision-language models. Recent audio work has begun to apply RL post-training to LALMs: Audio-Thinker~\citep{wu2025audiothinkerguidingaudiolanguage} uses GRPO to guide audio reasoning while R1-AQA~\citep{li2025reinforcementlearningoutperformssupervised} explores RL post-training for audio QA. However, these efforts target reasoning or classification accuracy and optimize text-level metrics. We are the first to apply GRPO with temporal verifiable rewards across a unified multi-task LALM.
\section{Tasks and Datasets}
\label{sec:tasks}

We train \textsc{Tempo} on five timestamping tasks spanning speech, sound,
and music using a combination of real and synthetic data. Table~\ref{tab:data_examples}
shows example inputs and outputs for each task.

\subsection{Real Data}

\paragraph{Multi-Speaker ASR and Diarization.} We use AMI~\citep{carletta2005ami},
ICSI~\citep{shriberg-etal-2004-icsi}, and Switchboard~\citep{godfrey1992switchboard},
three conversational corpora with speaker-labeled transcripts and precise timestamps,
enabling evaluation of both joint transcript-timing prediction and speaker turn
segmentation, yielding 16,045 ASR and 16,053 diarization samples.

\paragraph{Audio Temporal Grounding.} We use AudioSet Strong~\citep{hershey2021benefittemporallystronglabelsaudio}, a strongly-labeled subset of AudioSet~\citep{7952261} that provides onset and offset times for over 500 sound classes. To convert the categorical event labels into natural-language queries, we prompt GPT-4o to generate paraphrased queries for each class, yielding 13{,}500 training samples.

\paragraph{Dense Audio Captioning.} We directly use
TACOS~\citep{primus2025tacostemporallyalignedaudiocaptions}, which pairs audio
recordings with captions and onset/offset timestamps, yielding 9,858 data points for this task.

\paragraph{Timestamped Music Captioning.} We use
Slakh2100~\citep{manilow2019cuttingmusicsourceseparation}, a collection of 2,100
multi-track songs synthesized from MIDI, from which we derive timestamped chord, tempo,
and instrument annotations via the aligned MIDI files, yielding 13,000 samples. In total, the real-world corpus contains 68,456 training examples.

\begin{figure*}[t]
  \centering
  \includegraphics[width=\textwidth]{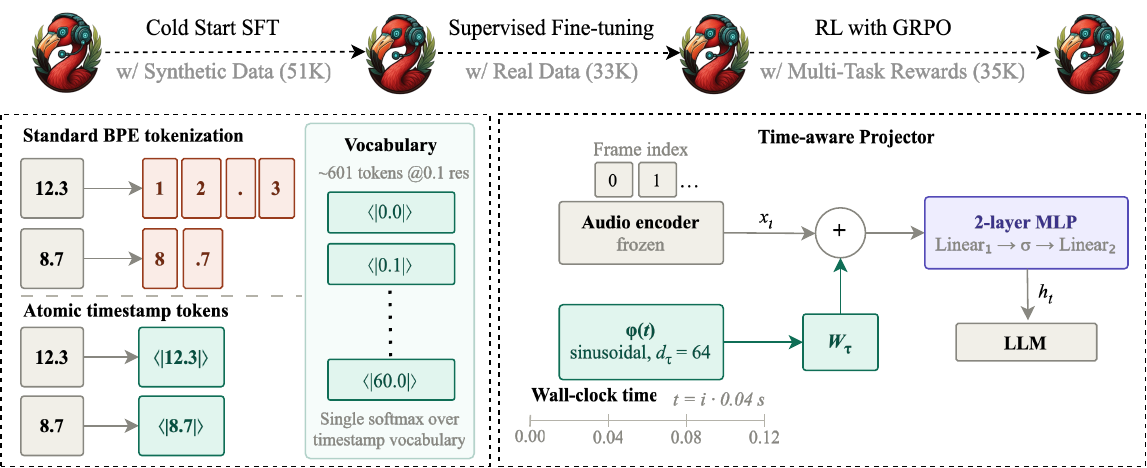}
  \vspace{-6mm}
  \caption{\small Overview of \textsc{Tempo}, built on Audio Flamingo 3. \textbf{Top:} Three-stage curriculum from synthetic SFT (51K) through real-data SFT (33K) to GRPO with verifiable rewards (35K). \textbf{Bottom-left:} Atomic timestamp tokens replace BPE fragmentation with ${\sim}601$ dedicated tokens at 0.1\,s resolution. \textbf{Bottom-right:} Time-aware projector injects sinusoidal wall-clock encodings $\varphi(t)$ into frozen audio encoder features before the MLP.}
  \label{fig:curriculum}
  \vspace{-2mm}
\end{figure*}

\subsection{Synthetic Data}

To improve temporal calibration before real-data training, we also construct a synthetic
Stage-1 corpus from LibriSpeech~\citep{7178964} and
ESC-50~\citep{piczak2015dataset} covering ASR, diarization, DAC, and AG tasks, with
timestamp tokens balanced uniformly across all possible values. We generate 8,783 ASR samples, 8,783 diarization samples, 11,962 AG samples and 11,984 DAC samples. We add 10,000 samples from Slakh2100 to this stage to ensure exposure to all tasks. This yields 51,512 training examples in total. Full details are in Appendix~\ref{app:datasets}.
\section{Method}

\textsc{Tempo} is built on Audio Flamingo 3 (AF3)~\citep{goel2025audioflamingo3advancing}, which couples a Whisper-large audio encoder to a Qwen2-7B language model through a two-layer MLP multi-modal projector. The audio encoder is frozen throughout training, and the five tasks are distinguished by task-specific tags in the prompt: \texttt{[speech:asr]}, \texttt{[speech:diar]}, \texttt{[audio:caption]}, \texttt{[audio:ground]}, and \texttt{[audio:music]}. Figure~\ref{fig:curriculum} gives an overview of TEMPO's architecture and training pipeline.

\subsection{Supervised Fine-Tuning}

We begin with a supervised fine-tuning (SFT) stage that teaches the model the timestamped output format and provides the temporal calibration needed before reinforcement learning. A standard teacher-forced cross-entropy objective is insufficient for fine-grained timestamping for three reasons. First, numeric timestamps rendered as text are tokenized inconsistently by the BPE vocabulary, fragmenting supervision across subword pieces. Second, the AF3 audio encoder produces features indexed by frame rather than wall-clock time, requiring the model to learn the correspondence implicitly. Third, token-level cross-entropy penalizes all incorrect timestamps equally, which is misaligned with the downstream temporal IoU (tIoU) metric. Our SFT recipe addresses each of these in turn through a corresponding architectural or objective change: atomic timestamp tokens, a time-aware multi-modal projector, and a distance-aware Gaussian loss.

\subsubsection{Atomic Timestamp Tokens}

To prevent BPE from fragmenting timestamps across subword pieces, we extend the tokenizer with atomic timestamp tokens of the form $\langle|t|\rangle$, where $t \in \{0.0, 0.1, \ldots, 60.0\}$ at $0.1$~s resolution. Each timestamp is represented by a single token, so a timestamp prediction becomes a single categorical decision over approximately 600 candidates. New token embeddings are initialized as the mean of the BPE decomposition of the corresponding numeric value.

\subsubsection{Time-Aware Multi-Modal Projector}

To supply the wall-clock information that frame-indexed features lack, we replace the AF3 projector, whose positional embeddings encode frame index rather than elapsed time, with one that injects sinusoidal wall-clock encodings into each frame embedding:
\begin{equation}
\mathbf{h}_t = \mathrm{Linear}_2\!\bigl(\sigma(\mathrm{Linear}_1(\mathbf{x}_t + W_\tau \phi(t)))\bigr),
\end{equation}
where $\mathbf{x}_t$ is the $t$-th audio frame, $\phi(t)$ is a fixed sinusoidal encoding of wall-clock time $t = i \cdot 0.04$~s, and $W_\tau$ is a learned projection initialized from $\mathcal{N}(0, 0.01^2)$. We use $d_\tau = 64$ with frequencies logarithmically spaced between periods of $0.08$~s and $60$~s, so that a single projector can span both sub-second acoustic events and long-range musical structure without task-specific parameterization.

\begin{figure}[t]
  \centering
  \includegraphics[width=0.9\columnwidth]{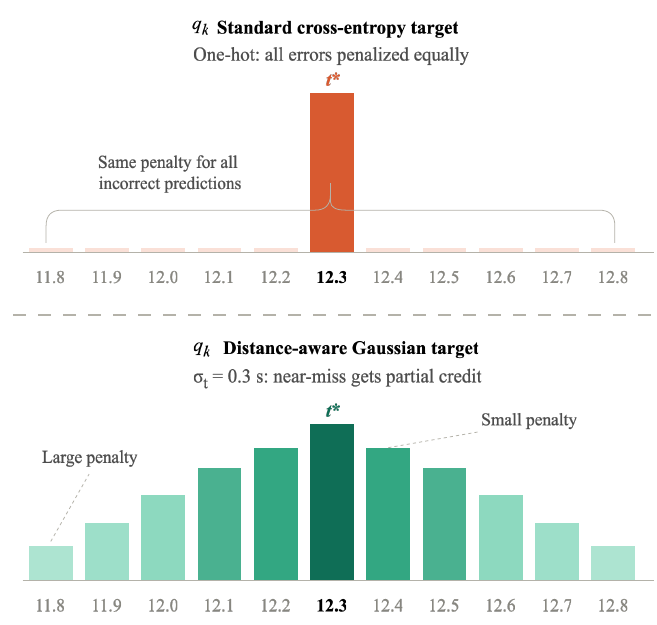}
    \caption{\small Standard cross-entropy (top) uses a one-hot target, penalizing all errors equally. Our distance-aware Gaussian target (bottom, $\sigma_t = 0.3$\,s) assigns partial credit to near-miss predictions, encouraging the model to learn the ordinal structure of the timestamp vocabulary.}
  \label{fig:dist-aware}
  \vspace{-5mm}
\end{figure}

\subsubsection{Distance-Aware Timestamp Loss}
\label{sec:distance_aware_loss}

Token-level cross-entropy treats every wrong timestamp as equally wrong, discarding the ordinal structure of the timestamp vocabulary and decoupling the training signal from the tIoU metric, under which a near miss is far better than a distant one. We restore this structure with an auxiliary loss that places a Gaussian soft label over the timestamp vocabulary, centered at the ground-truth time (Figure~\ref{fig:dist-aware}). If the target is $t^{\star}$ and the timestamp vocabulary corresponds to times $\{t_k\}_{k=1}^{K}$, the soft-label distribution is
\begin{equation}
q_k \propto \exp\!\left(-\frac{(t_k - t^{\star})^2}{2\sigma_t^2}\right), \qquad \sum_{k=1}^{K} q_k = 1.
\end{equation}
The auxiliary loss is
\begin{equation}
\mathcal{L}_{\mathrm{time}} = -\sum_k q_k \log p_k,
\end{equation}
where $p_k$ is the model probability over the timestamp vocabulary obtained by restricting the output logits to timestamp tokens and applying a softmax. The overall training objective is
\begin{equation}
\mathcal{L} = \mathcal{L}_{\mathrm{CE}} + \lambda_{\mathrm{time}} \mathcal{L}_{\mathrm{time}}
\end{equation}
with $\lambda_{\mathrm{time}} = 0.5$ and $\sigma_t = 0.3$~s, where $\mathcal{L}_{\mathrm{CE}}$ is the standard cross-entropy over all output positions including timestamps.

\subsubsection{Two-Stage Synthetic-to-Real Curriculum}

The three components above define \emph{what} the model learns to predict; the curriculum determines the order in which it sees supervision. Stage~1 trains on the $51{,}512$ synthetic examples described in Section~\ref{sec:tasks} to establish temporal calibration before exposure to real data. Stage~2 then fine-tunes the merged Stage~1 checkpoint on $32{,}726$ real-world examples taken from the real-world corpus.

\paragraph{Implementation details.} During SFT, LoRA adapters ($r{=}128$, $\alpha{=}256$, dropout $0.1$) are attached to the language-model attention and MLP projection matrices, while the multi-modal projector and timestamp token embeddings are trained directly. Stage~1 runs for two epochs at learning rate $1 \times 10^{-4}$ with 150 warmup steps; Stage~2 runs for two additional epochs at learning rate $5 \times 10^{-5}$ with 50 warmup steps. 

\subsection{Reinforcement Learning with GRPO}
\label{sec:rl}

Starting from the SFT checkpoint, we apply Group Relative Policy Optimization (GRPO)~\citep{Guo_2025} with verifiable rewards designed to directly optimize the temporal metrics used at evaluation. For each task, we construct a reward that averages the eval metrics themselves, so that maximizing reward and improving eval performance are aligned. All rewards are bounded in $[0, 1]$ and gated by a binary format check that returns zero if the output does not parse into the expected task tag with valid timestamp tokens.

\paragraph{ASR and diarization.} Both tasks use equally-weighted average of three eval metrics. For ASR, the reward averages $(1 - \mathrm{WER})$, symmetric mIoU between predicted and ground-truth segments, and a boundary score $\exp(-\mathrm{MAE}/\tau)$ (with $\tau = 1.0$~s) computed over greedy-matched pairs with IoU $>$ 0.3. For diarization, the same structure averages $(1 - \mathrm{DER})$, symmetric mIoU, and a soft speaker-labeled F1@IoU0.5 that requires exact speaker-label matches. Because both tasks frequently produce overlapping segments, we scale this averaged reward by a soft factor $(0.8 + 0.2 \cdot p)$, where $p$ is the fraction of non-overlapping output time.

\paragraph{Dense audio captioning.} The reward averages three equally-weighted terms: soft event F1@IoU0.5, symmetric mIoU, and a per-pair METEOR computed over greedy-matched predicted and ground-truth segments. This rewards both correct localization of sound events and accurate textual descriptions of each, jointly aligning the model with the three eval metrics rather than treating them as independent objectives.

\paragraph{Audio grounding.} The reward averages soft F1@IoU0.5, symmetric mIoU, and a boundary score with $\tau = 2.0$~s (larger than ASR's $\tau = 1.0$~s to accommodate the naturally noisier boundaries of environmental sound events). A light count-ratio term (weight $0.25$ out of $3.25$ total, or 7.7\%) discourages large mismatches between predicted and ground-truth interval counts. The weight is intentionally small because precision and recall in the F1 term already penalize over- and under-prediction; the count ratio acts as a soft regularizer rather than a hard gate.

\paragraph{Music captioning.} The reward is a weighted sum of five terms gated by a format check (weight $0.05$) that verifies all required section tags. The dominant component is chord accuracy (weight $0.45$): each ground-truth chord is scored against its best-overlapping prediction as $0.3 \cdot \mathrm{tIoU} + 0.7 \cdot \mathrm{sim}$, where similarity weights root pitch ($0.60$), quality ($0.25$), and extension ($0.15$), with a false-positive penalty $\bigl(0.7 + 0.3 \cdot \min(|G|/|P|,\,1)\bigr)$. Instrument accuracy (weight $0.20$) averages name matching with entry/exit timing error under a $2.0$\,s tolerance. Tempo accuracy (weight $0.15$) combines $0.7 \cdot \mathrm{BPM\;accuracy} + 0.3 \cdot \mathrm{timestamp\;accuracy}$ with the same tolerance. Statistics accuracy (weight $0.15$) averages relative note-density error and exact pitch-range matching. The chord boundary tolerance is relaxed to $1.0$\,s (from $0.5$\,s) to reduce reward sparsity early in training.

\paragraph{Implementation details.} GRPO training is initialized from the Stage~1+2 SFT checkpoint. We attach LoRA adapters ($r{=}256$, $\alpha{=}512$) to the language model and train for $1{,}000$ steps at learning rate $1 \times 10^{-5}$. Each step generates $8$ completions per prompt at temperature $1.2$, with a maximum completion length of $768$ tokens. The KL coefficient is set to $\beta = 0.01$.
\section{Experiments and Results}

We evaluate \textsc{Tempo} on held-out test sets across all five tasks, using metrics that capture both the textual and temporal aspects of each output.

\subsection{Evaluation Benchmarks}

We evaluate \textsc{Tempo} on held-out test sets across all five tasks. For dense audio captioning, we use the official TACOS test set: 2{,}000 question-answer pairs in which each query asks the model to caption all sound events in a clip along with their temporal boundaries. For audio temporal grounding, we reformat the TACOS test set into the grounding format used by TimeAudio~\citep{wang2025listeningframesbridgingtemporal}, in which each strong caption becomes a natural-language query and its associated segment becomes the ground-truth interval, yielding 5{,}151 question-answer pairs. For speaker diarization and multi-speaker ASR, we evaluate on the AMI Meeting test set, segmented into chunks of up to 60 seconds for consistency with our training setup; this yields 1{,}181 question-answer pairs per task, with the same audio chunks used for both and the expected output differing only in whether transcription is required. For timestamped music captioning, we evaluate on the Slakh2100 test set (225 tracks), split into 30-second segments to obtain 1{,}000 question-answer pairs, where each segment requires the model to produce timestamped chord labels, instrument entry and exit points, and tempo.

\subsection{Evaluation Metrics}

We report standard metrics for each task; full computational definitions are given in Appendix~\ref{app:eval_details}.

\paragraph{Speech metrics.} For multi-speaker ASR we report \emph{Word Error Rate (WER)}, computed over predicted and ground-truth segments matched by temporal IoU (Appendix~\ref{app:eval_details}). For diarization we report \emph{Diarization Error Rate (DER)} using the standard \texttt{pyannote.metrics} implementation~\citep{bredin17_interspeech} with a $0.25$~s collar; speaker labels are optimally permuted between hypothesis and reference before scoring, so a model that produces consistent but arbitrarily named speakers is not penalized for the naming alone.

\paragraph{Mean Intersection-over-Union (mIoU).} For all tasks that produce time intervals, we report a \emph{symmetric} mIoU computed as the harmonic mean of recall-side and precision-side mean-best-IoU. Unlike the more common recall-only mIoU, the symmetric formulation prevents over-prediction from inflating the score, since a model that emits many spurious intervals is penalized on the precision side.

\paragraph{F1 at IoU threshold and boundary MAE.} For dense audio captioning and audio grounding we report event \emph{F1@IoU0.5} (eF1@IoU), counting a predicted interval as a true positive if its IoU with a greedily-matched ground-truth interval is at least $0.5$. For diarization we additionally report a \emph{speaker-labeled F1@IoU0.5} (sF1@IoU) that requires exact speaker-label agreement in addition to temporal overlap. For tasks with continuous time boundaries (ASR, grounding, dense captioning, music) we report boundary \emph{MAE} in seconds, computed over greedy-matched pairs with $\mathrm{IoU} \geq 0.3$.

\paragraph{Caption and music metrics.} For dense audio captioning we additionally report \emph{METEOR} over matched segment pairs. For music captioning we report chord root and quality accuracy, chord F1@IoU0.5 and chord mIoU, instrument-name F1 and instrument boundary MAE, and tempo accuracy within a $4\%$ tolerance. Full definitions of all caption and music metrics are in Appendix~\ref{app:eval_details}.

\subsection{Compared Methods}
\label{sec:compared_methods}
 
We compare against three baseline families, frontier models, and ablations of our own pipeline. \textbf{Zero-shot LALMs} (Audio Flamingo 3, Audio Flamingo Next, and Qwen3-Omni) are evaluated without task-specific training. We additionally evaluate two \textbf{frontier models}, Gemini 2.5 Flash and Gemini 2.5 Pro~\citep{comanici2025gemini}, across all supported tasks. \textbf{TimeAudio}~\citep{wang2025listeningframesbridgingtemporal}, the closest prior open-source LALM-based timestamping approach, is evaluated on the two tasks it supports: dense captioning and audio grounding. \textbf{Naive SFT} follows our two-stage curriculum without the three proposed SFT innovations, isolating the effect of the training recipe from the data. \emph{Stage~1}, \emph{Stage~2}, and \emph{Stage~1+2} denote the SFT checkpoints obtained after the corresponding curriculum stages. Both \textbf{RL variants} are initialized from the Stage~1+2 checkpoint: \emph{single-task RL} performs a separate GRPO stage for each task using task-specific data and rewards, whereas \emph{multi-task RL} jointly optimizes all tasks in a single GRPO stage. We also compare against \textbf{specialized systems}, using one off-the-shelf dedicated model per task: Parakeet-TDT-0.6b-v2 \citep{xu2023efficientsequencetransductionjointly} for ASR, pyannote-3.1 \citep{bredin23_interspeech} for diarization, PretrainedSED/BEATs \citep{chen2022beatsaudiopretrainingacoustic} for dense captioning, CLAP-window \citep{wu2024largescalecontrastivelanguageaudiopretraining} for audio grounding, Chordino \citep{inproceedings} for chord estimation, and madmom \citep{böck2016madmomnewpythonaudio} for tempo estimation. All systems are evaluated on the same held-out test sets where applicable, and all generative-model results are obtained using deterministic greedy decoding.
 
\subsection{Main Results}
\begin{table*}[t]
  \centering
  \small
  \caption{Performance across all evaluation tasks. 
  \textbf{Speech:} WER: word error rate (\%); DER: diarization error rate (collar=0.25s) (\%). 
  \textbf{Sound:} eF1: event F1@IoU (\%); MET: METEOR (\%); F1: F1@IoU0.5 (\%). 
  \textbf{Music:} Root/Qual: chord root/quality accuracy (\%); Ch.F1: chord F1@IoU0.5 (\%); Inst F1: instrument name F1 (\%); Tempo: tempo accuracy within 4\% tolerance (\%). 
  \textbf{Shared:} MAE: boundary MAE (s, over matches with IoU$\geq$0.3); mIoU: symmetric mean IoU (\%); sF1: speaker-labeled F1@IoU (\%). 
  $\downarrow$: lower is better; $\uparrow$: higher is better. 
  $^{\dagger}$Specialist systems (off-the-shelf, one dedicated model per task): Parakeet-TDT-0.6b-v2 (ASR); pyannote-3.1 (diarization); PretrainedSED/BEATs (dense captioning); CLAP-window (audio grounding, instrument); Chordino (chord); madmom (tempo).}
  \label{tab:combined_results}

  \setlength{\tabcolsep}{3pt}
  \renewcommand{\arraystretch}{1.2}

  \resizebox{\textwidth}{!}{
  \begin{tabular}{l | ccc ccc | ccc ccc | cccc cc c}
  \toprule

  \multirow{3}{*}{\textbf{Model}}
    & \multicolumn{6}{c|}{\textbf{Speech}}
    & \multicolumn{6}{c|}{\textbf{Sound}}
    & \multicolumn{7}{c}{\textbf{Music}} \\

  \cmidrule(lr){2-7}
  \cmidrule(lr){8-13}
  \cmidrule(lr){14-20}

    & \multicolumn{3}{c}{\textbf{ASR}}
    & \multicolumn{3}{c|}{\textbf{Diarization}}
    & \multicolumn{3}{c}{\textbf{Dense Captioning}}
    & \multicolumn{3}{c|}{\textbf{Audio Grounding}}
    & \multicolumn{4}{c}{\textbf{Chord}}
    & \multicolumn{2}{c}{\textbf{Instrument}}
    & \textbf{Tempo} \\

  \cmidrule(lr){2-4}
  \cmidrule(lr){5-7}
  \cmidrule(lr){8-10}
  \cmidrule(lr){11-13}
  \cmidrule(lr){14-17}
  \cmidrule(lr){18-19}
  \cmidrule(lr){20-20}

    & MAE$\downarrow$ & mIoU$\uparrow$ & WER$\downarrow$
    & DER$\downarrow$ & mIoU$\uparrow$ & sF1$\uparrow$
    & eF1$\uparrow$ & mIoU$\uparrow$ & MET$\uparrow$
    & F1$\uparrow$  & MAE$\downarrow$ & mIoU$\uparrow$
    & Root$\uparrow$ & Qual$\uparrow$ & F1$\uparrow$ & mIoU$\uparrow$
    & F1$\uparrow$ & MAE$\downarrow$
    & Acc$\uparrow$ \\

  \midrule

  Audio Flamingo 3
    & 0.99 & 19.4 & 209.7
    & 103.2 & 2.9  & 1.0
    & 3.8  & 5.7  & 2.7
    & 3.3  & 1.25 & 5.6
    & 0.4 & 0.9 & 0.1 & 2.1
    & 33.7 & 2.84 & 18.0 \\

  Audio Flamingo Next
    & 1.39 & 11.5 & 167.1
    & 106.9 & 12.4 & 7.6
    & 11.6 & 22.3 & 11.9
    & 3.6  & 0.95 & 7.0
    & 0.4 & 1.2 & 0.0 & 3.2
    & 33.5 & 2.77 & 26.2 \\

  Qwen3-Omni
    & 1.12 & 31.6 & 69.7
    & 44.2 & 44.4 & 31.5
    & 57.9 & 65.4 & 19.0
    & 47.4 & 1.10 & 49.3
    & 1.1 & 2.8 & 0.5 & 6.9
    & 25.4 & 2.51 & 40.3 \\

  TimeAudio
    & -- & -- & --
    & -- & -- & --
    & 43.6 & 64.6 & 19.9
    & 40.1 & 1.32 & 46.3
    & -- & -- & -- & --
    & -- & -- & -- \\

  \midrule

  SFT (Naive) - Stage 1 + 2
    & 0.84 & 64.0 & 46.0
    & 27.4 & 67.2 & 52.3
    & 52.4 & 64.5 & 22.7
    & 37.1 & 1.44 & 40.3
    & 0.3 & 0.3 & 0.3 & 0.3
    & 0.5 & 3.95 & 37.8 \\

  \midrule

  SFT (Ours) - Stage 1
    & 2.42 & 41.7 & 94.0
    & 74.3 & 43.3 & 25.1
    & 47.6 & 50.9 & 10.5
    & 34.3 & 3.10 & 36.2
    & 19.4 & 32.4 & 5.8 & 75.5
    & 96.4 & 2.71 & 47.5 \\

  SFT (Ours) - Stage 2
    & 0.85 & 64.8 & 47.0
    & 25.1 & 70.0 & 54.5
    & 55.2 & 67.8 & 20.2
    & 44.6 & 1.14 & 44.7
    & 17.4 & 31.9 & 4.6 & 76.0
    & 89.4 & 2.78 & 41.7 \\

  SFT (Ours) - Stage 1 + 2
    & 0.77 & 63.8 & 44.7
    & 25.4 & 70.5 & 54.9
    & 58.5 & 68.3 & 22.3
    & 46.2 & 1.19 & 48.5
    & 19.6 & 32.7 & 6.2 & 76.2
    & 96.3 & 2.63 & 47.3 \\

  \midrule

  SFT + Single-task RL
    & 0.84 & 66.1 & 46.7
    & 26.8 & 69.2 & 54.0
    & 58.6 & 67.8 & 24.4
    & 45.5 & 1.21 & 48.0
    & 19.0 & 32.2 & 6.0 & 75.5
    & 96.3 & 2.63 & 47.0 \\

  \midrule

  SFT + Multi-task RL
    & 0.76 & 65.8 & 43.5
    & 25.4 & 71.1 & 56.3
    & 59.3 & 68.5 & 23.5
    & 46.5 & 1.32 & 49.4
    & 18.7 & 31.7 & 5.5 & 75.3
    & 96.0 & 2.62 & 46.9 \\

  \midrule

  Gemini 2.5 Flash
    & 1.01 & 43.6 & 49.4
    & 50.2 & 34.8 & 21.6
    & 21.5 & 52.1 & 10.7
    & 27.4 & 1.20 & 42.5
    & 8.6 & 19.2 & 1.1 & 30.4
    & 65.3 & 3.41 & 11.2 \\

  Gemini 2.5 Pro
    & 0.99 & 42.2 & 46.9
    & 27.6 & 51.7 & 42.0
    & 41.5 & 63.4 & 10.7
    & 36.9 & 1.09 & 48.7
    & 8.8 & 15.5 & 0.8 & 25.0
    & 77.4 & 3.03 & 12.9 \\

  \midrule

  Specialized systems$^{\dagger}$
    & 1.27 & 42.7 & 31.4
    & 19.6 & 55.3 & 48.8
    & 35.0 & 48.8 & 3.9
    & 30.8 & 1.83 & 43.6
    & 18.8 & 23.4 & 2.0 & 23.8
    & 61.2 & -- & 45.0 \\

  \bottomrule
  \end{tabular}
  }
\end{table*}
\paragraph{Multi-speaker ASR.} Existing LALMs perform poorly on multi-speaker ASR with timestamps under zero-shot evaluation. Qwen3-Omni performs best among baseline LALMs (WER of 69.7\%) but still fails to align transcripts to accurate temporal boundaries (MAE of 1.12~s, mIoU of 31.6\%). \textsc{Tempo} with full SFT (Stage~1+2) achieves a WER of 44.7\%, a 25-point absolute improvement over Qwen3-Omni, alongside the boundary MAE of 0.77~s and mIoU of 63.8\%. Adding multi-task RL on top of the SFT checkpoint yields further improvements: WER drops to 43.5\%, MAE to 0.76~s, and mIoU rises to 65.8\%.
 
\paragraph{Speaker diarization.} Qwen3-Omni achieves a reasonable DER of 44.2\% but speaker-labeled F1 (\textbf{sF1}) remains low at 31.5\%. \textsc{Tempo} reduces DER to 25.4\% and raises speaker-labeled F1 to 54.9\% and mIoU to 70.5\%. The Stage~1+2 curriculum provides a clear gain over Stage~2 alone (54.9\% vs 54.5\% sF1, 70.5\% vs 70.0\% mIoU), validating the synthetic-to-real curriculum design. Multi-task RL further improves sF1 to 56.3\% and mIoU to 71.1\% at the same DER.
 
\paragraph{Dense audio captioning.} \textsc{Tempo} surpasses Qwen3-Omni on both metrics with 58.5\% eF1 and 68.3\% mIoU, while matching it closely on METEOR (22.3\% vs 19.0\%). TimeAudio, the closest prior LALM-based approach, achieves 43.6\% eF1 and 64.6\% mIoU on the same evaluation, which \textsc{Tempo} exceeds by 14.9 and 3.7 absolute points respectively. Multi-task RL pushes performance further to 59.3\% eF1 and 68.5\% mIoU.
 
\paragraph{Audio temporal grounding.} Audio grounding shows the largest gap between zero-shot baselines and trained models. Audio Flamingo 3 and Audio Flamingo Next achieve F1@IoU0.5 of only 3.3\% and 3.6\%, respectively, indicating that these models lack the ability to localize natural-language queries to specific time intervals. Qwen3-Omni performs better at 47.4\% F1 but boundary MAE remains high at 1.10~s. \textsc{Tempo} with full SFT reaches 46.2\% F1 with boundary MAE of 1.19~s and mIoU of 48.5\%. TimeAudio achieves 40.1\% F1 and 46.3\% mIoU on the same eval, which \textsc{Tempo} exceeds by 6.1 and 2.2 absolute points. Multi-task RL improves F1 to 46.5\% and mIoU to 49.4\%.
 
\paragraph{Timestamped music captioning.} On Slakh2100, the zero-shot baselines fail almost completely on chord-related metrics: all three models produce chord F1@IoU below 1\% and chord root accuracy below 2\%, indicating that none has learned to associate harmonic events with precise timestamps. Tempo accuracy is also poor for the LALM baselines (18.0\% for Audio Flamingo 3, 40.3\% for Qwen3-Omni). \textsc{Tempo} substantially closes these gaps: Stage~1+2 reaches 19.6\% chord root accuracy, 32.7\% chord quality accuracy, 6.2\% chord F1@IoU0.5, and 76.2\% chord mIoU, alongside 96.3\% instrument F1 and 47.3\% tempo accuracy.
 
\subsection{Ablation Study}
 
\paragraph{Effect of the synthetic-to-real curriculum.} Comparing Stage~1 alone and Stage~1+2 reveals the contribution of the two-stage curriculum. Stage~1 trained on synthetic data alone is insufficient for downstream tasks (94.0\% WER for ASR, 74.3\% DER for diarization, 34.3\% F1 for grounding). Stage~2 fine-tuning on real annotated data brings dramatic improvements across the board, with the largest gains on audio grounding (F1 from 34.3\% to 46.2\%) and ASR (WER from 94.0\% to 44.7\%). Stage~1+2 also outperforms Stage~2 alone in most cases (0.77~s vs 0.85~s ASR MAE, 70.5\% vs 70.0\% diarization mIoU), indicating that the synthetic warmup provides priors that transfer beneficially to real data.

\paragraph{Effect of the full SFT recipe.} On the full Stage~1+2 corpus, the Naive SFT baseline isolates the contribution of our three SFT innovations under the complete training budget. The full SFT recipe outperforms Naive SFT: 1.3 points lower WER on ASR, 2.0 points lower DER on diarization, +6.1 event F1 on dense captioning, and +9.1 F1 on audio grounding. A controlled study isolating each component individually is provided in Appendix~\ref{app:component_ablation}.

\paragraph{Effect of reinforcement learning.} Adding GRPO on top of the Stage~1+2 SFT checkpoint produces consistent but moderate improvements across 4 out of the 5 tasks (Table~\ref{tab:combined_results}). Multi-task RL improves ASR WER by 1.2 points (44.7 to 43.5), diarization sF1 by 1.4 points (54.9 to 56.3), dense captioning eF1 by 0.8 points (58.5 to 59.3), and audio grounding mIoU by 0.9 points (48.5 to 49.4). Single-task RL, in which a separate RL stage is run for each task, yields gains of similar magnitude but does not consistently match multi-task RL across all metrics. However, RL did not help on music captioning, where nearly every metric declines slightly relative to the SFT checkpoint. We hypothesize that this lack of improvement on music is because of the lack of pretrained music data in the AF3's training curriculum. The smaller magnitude of these RL gains relative to the SFT gains supports our overall finding: careful SFT design is the primary lever for installing timestamping in LALMs, with RL providing additional refinement of timestamp boundaries.

\section{Baseline format-failure analysis}
To ensure that the large gaps to zero-shot LALMs are not artifacts of output formatting, we use model-specific tolerant parsers and separately evaluate each baseline on only its parseable outputs. The performance gap remains substantial even after removing format failures, indicating that the dominant errors arise from incorrect content and temporal localization rather than parsing; detailed prompts, valid-output analysis, and qualitative failure modes are provided in Appendix \ref{app:baseline_failure_analysis}.   Representative qualitative examples in Appendix~\ref{app:qualitative_analysis}
show that TEMPO's residual errors are dominated by segment merging and
boundary misalignment, even when the underlying semantic content is
correct.

\section{Transfer to real world music}
All our training data consist of human annotated labels from real-world recordings except music. Slakh2100 \citep{manilow2019cuttingmusicsourceseparation} dataset comprises of human annotated MIDI files from Lakh Midi \citep{raffel2016learning} but the recordings are synthesized from the MIDI files. To evaluate whether training on synthetic data transfers to real world recordings, we evaluate our trained model on Maestro dataset \citep{hawthorne2018enabling} which consists of real-world piano recordings. We use the same data processing pipeline from the testset of the Maestro dataset. We randomly pick 1000 samples to evaluate our final checkpoint. The final numbers are in Table \ref{tab:maestro_results}. Our multi-task RL-trained model consistently outperforms the base
Audio Flamingo 3 model and Gemini 2.5 Flash across all metrics.

\begin{table}[t]
  \centering
  \small
  \caption{
  Out-of-domain music evaluation on real acoustic piano recordings
  from MAESTRO.  
  }
  \label{tab:maestro_results}
  \setlength{\tabcolsep}{5pt}
  \renewcommand{\arraystretch}{1.15}

  \begin{tabular}{l cccc}
    \toprule
    \textbf{Model}
      & \textbf{Root}$\uparrow$
      & \textbf{Qual}$\uparrow$
      & \textbf{Inst F1}$\uparrow$
      & \textbf{Tempo}$\uparrow$ \\
    \midrule

    AF3
      & 0.0
      & 0.0
      & 98.7
      & 19.5 \\

    Gemini 2.5 Flash
      & 9.1
      & 22.4
      & 96.3
      & 17.9 \\

    \textbf{TEMPO}
      & \textbf{23.9}
      & \textbf{27.3}
      & \textbf{99.8}
      & \textbf{24.8} \\

    \bottomrule
  \end{tabular}
\end{table}
\section{Conclusion}

We presented \textsc{Tempo}, the first unified LALM for timestamped generation across speech, sound, and music, covering multi-speaker ASR, speaker diarization, audio temporal grounding, dense audio captioning, and timestamped music captioning in a single decoder. Our SFT recipe combines atomic timestamp tokens, a time-aware multi-modal projector, and a distance-aware Gaussian loss, trained under a synthetic-to-real curriculum and refined with GRPO against verifiable temporal rewards. \textsc{Tempo} substantially outperforms Audio Flamingo Next and Qwen3-Omni, while remaining competitive with Gemini 2.5 models on a benchmark spanning five tasks, with the largest gains on multi-speaker ASR (WER from 69.7\% to 43.5\%) and diarization (mIoU from 44.4\% to 71.1\%). Ablations show that careful SFT design is the primary lever for installing timestamping in LALMs, with RL providing additional refinement.
\newpage
\section*{Limitations}
TEMPO covers five timestamping tasks but does not address all forms of temporal audio understanding. In particular, we do not handle overlapping multi-domain events (e.g., simultaneous music, background sounds, and speech), which would require the model to jointly segment and label multiple concurrent streams.

Second, TEMPO uses a fixed temporal resolution of 0.1\,s. This resolution is too coarse for applications requiring sub-100\,ms precision, such as phoneme-level alignment or fine-grained percussive onset detection, which are outside the scope of this work. For the five tasks considered here, however, timestamp quantization is not the dominant source of error: a 0.1\,s grid introduces at most 0.05\,s of quantization error, whereas the observed boundary MAE is approximately 0.77 - 1.19\,s for the corresponding temporal tasks. Diarization is additionally evaluated with a 0.25\,s collar. Thus, finer timestamp resolution is unlikely to materially affect the reported results, although adaptive or higher-resolution timestamp representations may be necessary for tasks requiring substantially finer temporal precision.

Third, TEMPO is built on Audio Flamingo 3 with a frozen Whisper-large encoder, inheriting any limitations of that encoder's spectral and temporal representations. Our results are thus conditioned on this specific encoder and may not generalize to other audio frontends.

Fourth, we do not perform multi-seed training due to the computational cost of repeating the full multi-stage 7B training pipeline; however, the large margins over the strongest baselines and the consistent ordering across our controlled ablations (Appendix~\ref{app:component_ablation}) support the robustness of our main conclusions.

Finally, the music captioning component is trained exclusively on Slakh2100, a synthesized MIDI dataset. We additionally evaluate the trained checkpoint on the real-world MAESTRO dataset, demonstrating transfer beyond synthesized MIDI recordings. However, performance on broader real-world music with natural acoustics, mixing artifacts, and production effects may differ substantially.

\bibliography{custom}

@misc{ghosh2026audioflamingonextnextgeneration,
      title={Audio Flamingo Next: Next-Generation Open Audio-Language Models for Speech, Sound, and Music}, 
      author={Sreyan Ghosh and Arushi Goel and Kaousheik Jayakumar and Lasha Koroshinadze and Nishit Anand and Zhifeng Kong and Siddharth Gururani and Sang-gil Lee and Jaehyeon Kim and Aya Aljafari and Chao-Han Huck Yang and Sungwon Kim and Ramani Duraiswami and Dinesh Manocha and Mohammad Shoeybi and Bryan Catanzaro and Ming-Yu Liu and Wei Ping},
      year={2026},
      eprint={2604.10905},
      archivePrefix={arXiv},
      primaryClass={cs.SD},
      url={https://arxiv.org/abs/2604.10905}, 
}

@misc{goel2025audioflamingo3advancing,
      title={Audio Flamingo 3: Advancing Audio Intelligence with Fully Open Large Audio Language Models}, 
      author={Arushi Goel and Sreyan Ghosh and Jaehyeon Kim and Sonal Kumar and Zhifeng Kong and Sang-gil Lee and Chao-Han Huck Yang and Ramani Duraiswami and Dinesh Manocha and Rafael Valle and Bryan Catanzaro},
      year={2025},
      eprint={2507.08128},
      archivePrefix={arXiv},
      primaryClass={cs.SD},
      url={https://arxiv.org/abs/2507.08128}, 
}

@misc{kumar2026tactimestampedaudiocaptioning,
      title={TAC: Timestamped Audio Captioning}, 
      author={Sonal Kumar and Prem Seetharaman and Ke Chen and Oriol Nieto and Jiaqi Su and Zhepei Wang and Rithesh Kumar and Dinesh Manocha and Nicholas J. Bryan and Zeyu Jin and Justin Salamon},
      year={2026},
      eprint={2602.15766},
      archivePrefix={arXiv},
      primaryClass={cs.SD},
      url={https://arxiv.org/abs/2602.15766}, 
}

@misc{ghosh2025musicflamingoscalingmusic,
      title={Music Flamingo: Scaling Music Understanding in Audio Language Models}, 
      author={Sreyan Ghosh and Arushi Goel and Lasha Koroshinadze and Sang-gil Lee and Zhifeng Kong and Joao Felipe Santos and Ramani Duraiswami and Dinesh Manocha and Wei Ping and Mohammad Shoeybi and Bryan Catanzaro},
      year={2025},
      eprint={2511.10289},
      archivePrefix={arXiv},
      primaryClass={eess.AS},
      url={https://arxiv.org/abs/2511.10289}, 
}

@misc{ghosh2024gamalargeaudiolanguagemodel,
      title={GAMA: A Large Audio-Language Model with Advanced Audio Understanding and Complex Reasoning Abilities}, 
      author={Sreyan Ghosh and Sonal Kumar and Ashish Seth and Chandra Kiran Reddy Evuru and Utkarsh Tyagi and S Sakshi and Oriol Nieto and Ramani Duraiswami and Dinesh Manocha},
      year={2024},
      eprint={2406.11768},
      archivePrefix={arXiv},
      primaryClass={cs.SD},
      url={https://arxiv.org/abs/2406.11768}, 
}

@misc{kimiteam2025kimiaudiotechnicalreport,
      title={Kimi-Audio Technical Report}, 
      author={KimiTeam and Ding Ding and Zeqian Ju and Yichong Leng and Songxiang Liu and Tong Liu and Zeyu Shang and Kai Shen and Wei Song and Xu Tan and Heyi Tang and Zhengtao Wang and Chu Wei and Yifei Xin and Xinran Xu and Jianwei Yu and Yutao Zhang and Xinyu Zhou and Y. Charles and Jun Chen and Yanru Chen and Yulun Du and Weiran He and Zhenxing Hu and Guokun Lai and Qingcheng Li and Yangyang Liu and Weidong Sun and Jianzhou Wang and Yuzhi Wang and Yuefeng Wu and Yuxin Wu and Dongchao Yang and Hao Yang and Ying Yang and Zhilin Yang and Aoxiong Yin and Ruibin Yuan and Yutong Zhang and Zaida Zhou},
      year={2025},
      eprint={2504.18425},
      archivePrefix={arXiv},
      primaryClass={eess.AS},
      url={https://arxiv.org/abs/2504.18425}, 
}

@misc{ghosh2025audioflamingo2audiolanguage,
      title={Audio Flamingo 2: An Audio-Language Model with Long-Audio Understanding and Expert Reasoning Abilities}, 
      author={Sreyan Ghosh and Zhifeng Kong and Sonal Kumar and S Sakshi and Jaehyeon Kim and Wei Ping and Rafael Valle and Dinesh Manocha and Bryan Catanzaro},
      year={2025},
      eprint={2503.03983},
      archivePrefix={arXiv},
      primaryClass={cs.SD},
      url={https://arxiv.org/abs/2503.03983}, 
}

@misc{kong2024audioflamingonovelaudio,
      title={Audio Flamingo: A Novel Audio Language Model with Few-Shot Learning and Dialogue Abilities}, 
      author={Zhifeng Kong and Arushi Goel and Rohan Badlani and Wei Ping and Rafael Valle and Bryan Catanzaro},
      year={2024},
      eprint={2402.01831},
      archivePrefix={arXiv},
      primaryClass={cs.SD},
      url={https://arxiv.org/abs/2402.01831}, 
}

@article{Guo_2025,
   title={DeepSeek-R1 incentivizes reasoning in LLMs through reinforcement learning},
   volume={645},
   ISSN={1476-4687},
   url={http://dx.doi.org/10.1038/s41586-025-09422-z},
   DOI={10.1038/s41586-025-09422-z},
   number={8081},
   journal={Nature},
   publisher={Springer Science and Business Media LLC},
   author={Guo, Daya and Yang, Dejian and Zhang, Haowei and Song, Junxiao and Wang, Peiyi and Zhu, Qihao and Xu, Runxin and Zhang, Ruoyu and Ma, Shirong and Bi, Xiao and Zhang, Xiaokang and Yu, Xingkai and Wu, Yu and Wu, Z. F. and Gou, Zhibin and Shao, Zhihong and Li, Zhuoshu and Gao, Ziyi and Liu, Aixin and Xue, Bing and Wang, Bingxuan and Wu, Bochao and Feng, Bei and Lu, Chengda and Zhao, Chenggang and Deng, Chengqi and Ruan, Chong and Dai, Damai and Chen, Deli and Ji, Dongjie and Li, Erhang and Lin, Fangyun and Dai, Fucong and Luo, Fuli and Hao, Guangbo and Chen, Guanting and Li, Guowei and Zhang, H. and Xu, Hanwei and Ding, Honghui and Gao, Huazuo and Qu, Hui and Li, Hui and Guo, Jianzhong and Li, Jiashi and Chen, Jingchang and Yuan, Jingyang and Tu, Jinhao and Qiu, Junjie and Li, Junlong and Cai, J. L. and Ni, Jiaqi and Liang, Jian and Chen, Jin and Dong, Kai and Hu, Kai and You, Kaichao and Gao, Kaige and Guan, Kang and Huang, Kexin and Yu, Kuai and Wang, Lean and Zhang, Lecong and Zhao, Liang and Wang, Litong and Zhang, Liyue and Xu, Lei and Xia, Leyi and Zhang, Mingchuan and Zhang, Minghua and Tang, Minghui and Zhou, Mingxu and Li, Meng and Wang, Miaojun and Li, Mingming and Tian, Ning and Huang, Panpan and Zhang, Peng and Wang, Qiancheng and Chen, Qinyu and Du, Qiushi and Ge, Ruiqi and Zhang, Ruisong and Pan, Ruizhe and Wang, Runji and Chen, R. J. and Jin, R. L. and Chen, Ruyi and Lu, Shanghao and Zhou, Shangyan and Chen, Shanhuang and Ye, Shengfeng and Wang, Shiyu and Yu, Shuiping and Zhou, Shunfeng and Pan, Shuting and Li, S. S. and Zhou, Shuang and Wu, Shaoqing and Yun, Tao and Pei, Tian and Sun, Tianyu and Wang, T. and Zeng, Wangding and Liu, Wen and Liang, Wenfeng and Gao, Wenjun and Yu, Wenqin and Zhang, Wentao and Xiao, W. L. and An, Wei and Liu, Xiaodong and Wang, Xiaohan and Chen, Xiaokang and Nie, Xiaotao and Cheng, Xin and Liu, Xin and Xie, Xin and Liu, Xingchao and Yang, Xinyu and Li, Xinyuan and Su, Xuecheng and Lin, Xuheng and Li, X. Q. and Jin, Xiangyue and Shen, Xiaojin and Chen, Xiaosha and Sun, Xiaowen and Wang, Xiaoxiang and Song, Xinnan and Zhou, Xinyi and Wang, Xianzu and Shan, Xinxia and Li, Y. K. and Wang, Y. Q. and Wei, Y. X. and Zhang, Yang and Xu, Yanhong and Li, Yao and Zhao, Yao and Sun, Yaofeng and Wang, Yaohui and Yu, Yi and Zhang, Yichao and Shi, Yifan and Xiong, Yiliang and He, Ying and Piao, Yishi and Wang, Yisong and Tan, Yixuan and Ma, Yiyang and Liu, Yiyuan and Guo, Yongqiang and Ou, Yuan and Wang, Yuduan and Gong, Yue and Zou, Yuheng and He, Yujia and Xiong, Yunfan and Luo, Yuxiang and You, Yuxiang and Liu, Yuxuan and Zhou, Yuyang and Zhu, Y. X. and Huang, Yanping and Li, Yaohui and Zheng, Yi and Zhu, Yuchen and Ma, Yunxian and Tang, Ying and Zha, Yukun and Yan, Yuting and Ren, Z. Z. and Ren, Zehui and Sha, Zhangli and Fu, Zhe and Xu, Zhean and Xie, Zhenda and Zhang, Zhengyan and Hao, Zhewen and Ma, Zhicheng and Yan, Zhigang and Wu, Zhiyu and Gu, Zihui and Zhu, Zijia and Liu, Zijun and Li, Zilin and Xie, Ziwei and Song, Ziyang and Pan, Zizheng and Huang, Zhen and Xu, Zhipeng and Zhang, Zhongyu and Zhang, Zhen},
   year={2025},
   month=Sept, pages={633–638} }

@inproceedings{audiocaps,
  title={AudioCaps: Generating Captions for Audios in The Wild},
  author={Kim, Chris Dongjoo and Kim, Byeongchang and Lee, Hyunmin and Kim, Gunhee},
  booktitle={NAACL-HLT},
  year={2019}
}

@misc{drossos2019clothoaudiocaptioningdataset,
      title={Clotho: An Audio Captioning Dataset}, 
      author={Konstantinos Drossos and Samuel Lipping and Tuomas Virtanen},
      year={2019},
      eprint={1910.09387},
      archivePrefix={arXiv},
      primaryClass={cs.SD},
      url={https://arxiv.org/abs/1910.09387}, 
}

@misc{wang2025listeningframesbridgingtemporal,
      title={Listening Between the Frames: Bridging Temporal Gaps in Large Audio-Language Models}, 
      author={Hualei Wang and Yiming Li and Shuo Ma and Hong Liu and Xiangdong Wang},
      year={2025},
      eprint={2511.11039},
      archivePrefix={arXiv},
      primaryClass={cs.SD},
      url={https://arxiv.org/abs/2511.11039}, 
}

@misc{primus2025tacostemporallyalignedaudiocaptions,
      title={TACOS: Temporally-aligned Audio CaptiOnS for Language-Audio Pretraining}, 
      author={Paul Primus and Florian Schmid and Gerhard Widmer},
      year={2025},
      eprint={2505.07609},
      archivePrefix={arXiv},
      primaryClass={eess.AS},
      url={https://arxiv.org/abs/2505.07609}, 
}

@misc{manilow2019cuttingmusicsourceseparation,
      title={Cutting Music Source Separation Some Slakh: A Dataset to Study the Impact of Training Data Quality and Quantity}, 
      author={Ethan Manilow and Gordon Wichern and Prem Seetharaman and Jonathan Le Roux},
      year={2019},
      eprint={1909.08494},
      archivePrefix={arXiv},
      primaryClass={cs.SD},
      url={https://arxiv.org/abs/1909.08494}, 
}

@INPROCEEDINGS{7952261,
  author={Gemmeke, Jort F. and Ellis, Daniel P. W. and Freedman, Dylan and Jansen, Aren and Lawrence, Wade and Moore, R. Channing and Plakal, Manoj and Ritter, Marvin},
  booktitle={2017 IEEE International Conference on Acoustics, Speech and Signal Processing (ICASSP)}, 
  title={Audio Set: An ontology and human-labeled dataset for audio events}, 
  year={2017},
  volume={},
  number={},
  pages={776-780},
  doi={10.1109/ICASSP.2017.7952261}
  }

@misc{openai2024gpt4technicalreport,
      title={GPT-4 Technical Report}, 
      author={OpenAI and Josh Achiam and Steven Adler and Sandhini Agarwal and Lama Ahmad and Ilge Akkaya and Florencia Leoni Aleman and Diogo Almeida and Janko Altenschmidt and Sam Altman and Shyamal Anadkat and Red Avila and Igor Babuschkin and Suchir Balaji and Valerie Balcom and Paul Baltescu and Haiming Bao and Mohammad Bavarian and Jeff Belgum and Irwan Bello and Jake Berdine and Gabriel Bernadett-Shapiro and Christopher Berner and Lenny Bogdonoff and Oleg Boiko and Madelaine Boyd and Anna-Luisa Brakman and Greg Brockman and Tim Brooks and Miles Brundage and Kevin Button and Trevor Cai and Rosie Campbell and Andrew Cann and Brittany Carey and Chelsea Carlson and Rory Carmichael and Brooke Chan and Che Chang and Fotis Chantzis and Derek Chen and Sully Chen and Ruby Chen and Jason Chen and Mark Chen and Ben Chess and Chester Cho and Casey Chu and Hyung Won Chung and Dave Cummings and Jeremiah Currier and Yunxing Dai and Cory Decareaux and Thomas Degry and Noah Deutsch and Damien Deville and Arka Dhar and David Dohan and Steve Dowling and Sheila Dunning and Adrien Ecoffet and Atty Eleti and Tyna Eloundou and David Farhi and Liam Fedus and Niko Felix and Simón Posada Fishman and Juston Forte and Isabella Fulford and Leo Gao and Elie Georges and Christian Gibson and Vik Goel and Tarun Gogineni and Gabriel Goh and Rapha Gontijo-Lopes and Jonathan Gordon and Morgan Grafstein and Scott Gray and Ryan Greene and Joshua Gross and Shixiang Shane Gu and Yufei Guo and Chris Hallacy and Jesse Han and Jeff Harris and Yuchen He and Mike Heaton and Johannes Heidecke and Chris Hesse and Alan Hickey and Wade Hickey and Peter Hoeschele and Brandon Houghton and Kenny Hsu and Shengli Hu and Xin Hu and Joost Huizinga and Shantanu Jain and Shawn Jain and Joanne Jang and Angela Jiang and Roger Jiang and Haozhun Jin and Denny Jin and Shino Jomoto and Billie Jonn and Heewoo Jun and Tomer Kaftan and Łukasz Kaiser and Ali Kamali and Ingmar Kanitscheider and Nitish Shirish Keskar and Tabarak Khan and Logan Kilpatrick and Jong Wook Kim and Christina Kim and Yongjik Kim and Jan Hendrik Kirchner and Jamie Kiros and Matt Knight and Daniel Kokotajlo and Łukasz Kondraciuk and Andrew Kondrich and Aris Konstantinidis and Kyle Kosic and Gretchen Krueger and Vishal Kuo and Michael Lampe and Ikai Lan and Teddy Lee and Jan Leike and Jade Leung and Daniel Levy and Chak Ming Li and Rachel Lim and Molly Lin and Stephanie Lin and Mateusz Litwin and Theresa Lopez and Ryan Lowe and Patricia Lue and Anna Makanju and Kim Malfacini and Sam Manning and Todor Markov and Yaniv Markovski and Bianca Martin and Katie Mayer and Andrew Mayne and Bob McGrew and Scott Mayer McKinney and Christine McLeavey and Paul McMillan and Jake McNeil and David Medina and Aalok Mehta and Jacob Menick and Luke Metz and Andrey Mishchenko and Pamela Mishkin and Vinnie Monaco and Evan Morikawa and Daniel Mossing and Tong Mu and Mira Murati and Oleg Murk and David Mély and Ashvin Nair and Reiichiro Nakano and Rajeev Nayak and Arvind Neelakantan and Richard Ngo and Hyeonwoo Noh and Long Ouyang and Cullen O'Keefe and Jakub Pachocki and Alex Paino and Joe Palermo and Ashley Pantuliano and Giambattista Parascandolo and Joel Parish and Emy Parparita and Alex Passos and Mikhail Pavlov and Andrew Peng and Adam Perelman and Filipe de Avila Belbute Peres and Michael Petrov and Henrique Ponde de Oliveira Pinto and Michael and Pokorny and Michelle Pokrass and Vitchyr H. Pong and Tolly Powell and Alethea Power and Boris Power and Elizabeth Proehl and Raul Puri and Alec Radford and Jack Rae and Aditya Ramesh and Cameron Raymond and Francis Real and Kendra Rimbach and Carl Ross and Bob Rotsted and Henri Roussez and Nick Ryder and Mario Saltarelli and Ted Sanders and Shibani Santurkar and Girish Sastry and Heather Schmidt and David Schnurr and John Schulman and Daniel Selsam and Kyla Sheppard and Toki Sherbakov and Jessica Shieh and Sarah Shoker and Pranav Shyam and Szymon Sidor and Eric Sigler and Maddie Simens and Jordan Sitkin and Katarina Slama and Ian Sohl and Benjamin Sokolowsky and Yang Song and Natalie Staudacher and Felipe Petroski Such and Natalie Summers and Ilya Sutskever and Jie Tang and Nikolas Tezak and Madeleine B. Thompson and Phil Tillet and Amin Tootoonchian and Elizabeth Tseng and Preston Tuggle and Nick Turley and Jerry Tworek and Juan Felipe Cerón Uribe and Andrea Vallone and Arun Vijayvergiya and Chelsea Voss and Carroll Wainwright and Justin Jay Wang and Alvin Wang and Ben Wang and Jonathan Ward and Jason Wei and CJ Weinmann and Akila Welihinda and Peter Welinder and Jiayi Weng and Lilian Weng and Matt Wiethoff and Dave Willner and Clemens Winter and Samuel Wolrich and Hannah Wong and Lauren Workman and Sherwin Wu and Jeff Wu and Michael Wu and Kai Xiao and Tao Xu and Sarah Yoo and Kevin Yu and Qiming Yuan and Wojciech Zaremba and Rowan Zellers and Chong Zhang and Marvin Zhang and Shengjia Zhao and Tianhao Zheng and Juntang Zhuang and William Zhuk and Barret Zoph},
      year={2024},
      eprint={2303.08774},
      archivePrefix={arXiv},
      primaryClass={cs.CL},
      url={https://arxiv.org/abs/2303.08774}, 
}

@inproceedings{shriberg-etal-2004-icsi,
    title = "The {ICSI} Meeting Recorder Dialog Act ({MRDA}) Corpus",
    author = "Shriberg, Elizabeth  and
      Dhillon, Raj  and
      Bhagat, Sonali  and
      Ang, Jeremy  and
      Carvey, Hannah",
    booktitle = "Proceedings of the 5th {SIG}dial Workshop on Discourse and Dialogue at {HLT}-{NAACL} 2004",
    month = apr # " 30 - " # may # " 1",
    year = "2004",
    address = "Cambridge, Massachusetts, USA",
    publisher = "Association for Computational Linguistics",
    url = "https://aclanthology.org/W04-2319/",
    pages = "97--100"
}

@inproceedings{godfrey1992switchboard,
  title={SWITCHBOARD: Telephone Speech Corpus for Research and Development},
  author={Godfrey, John J. and Holliman, Edward C. and McDaniel, Jane},
  booktitle={Proceedings of the IEEE International Conference on Acoustics, Speech and Signal Processing (ICASSP)},
  volume={1},
  pages={517--520},
  year={1992},
  publisher={IEEE},
  doi={10.1109/ICASSP.1992.225858}
}

@inproceedings{carletta2005ami,
  title={The {AMI} Meeting Corpus: A Pre-announcement},
  author={Carletta, Jean and Ashby, Simone and Bourban, Sebastien and Flynn, Mike and Guillemot, Ma{\"e}l and Hain, Thomas and Kadlec, Jaroslav and Karaiskos, Vasilis and Kraaij, Wessel and Kronenthal, Melissa and Lathoud, Guillaume and Lincoln, Mike and Lisowska, Agn{\`e}s and McCowan, Iain and Post, Wilfried and Reidsma, Dennis and Wellner, Pierre},
  booktitle={Machine Learning for Multimodal Interaction (MLMI)},
  series={Lecture Notes in Computer Science},
  volume={3869},
  pages={28--39},
  year={2005},
  publisher={Springer},
  doi={10.1007/11677482_3}
}

@misc{hershey2021benefittemporallystronglabelsaudio,
      title={The Benefit Of Temporally-Strong Labels In Audio Event Classification}, 
      author={Shawn Hershey and Daniel P W Ellis and Eduardo Fonseca and Aren Jansen and Caroline Liu and R Channing Moore and Manoj Plakal},
      year={2021},
      eprint={2105.07031},
      archivePrefix={arXiv},
      primaryClass={cs.SD},
      url={https://arxiv.org/abs/2105.07031}, 
}

@inproceedings{bredin17_interspeech,
  title     = {{ pyannote.metrics: A Toolkit for Reproducible Evaluation, Diagnostic, and Error Analysis of Speaker Diarization Systems}},
  author    = {Hervé Bredin},
  year      = {2017},
  booktitle = {{Interspeech 2017}},
  pages     = {3587--3591},
  doi       = {10.21437/Interspeech.2017-411},
  issn      = {2958-1796},
}

@inproceedings{burgoyne2011expert,
  title={An Expert Ground Truth Set for Audio Chord 
         Recognition and Music Analysis},
  author={Burgoyne, John Ashley and Wild, Jonathan 
          and Fujinaga, Ichiro},
  booktitle={ISMIR},
  year={2011}
}

@inproceedings{diarizationLM, series={interspeech\_2024},
   title={DiarizationLM: Speaker Diarization Post-Processing with Large Language Models},
   url={http://dx.doi.org/10.21437/Interspeech.2024-209},
   DOI={10.21437/interspeech.2024-209},
   booktitle={Interspeech 2024},
   publisher={ISCA},
   author={Wang, Quan and Huang, Yiling and Zhao, Guanlong and Clark, Evan and Xia, Wei and Liao, Hank},
   year={2024},
   month=Sept, pages={3754–3758},
   collection={interspeech_2024} }

@inproceedings{
  hawthorne2018enabling,
  title={Enabling Factorized Piano Music Modeling and Generation with the {MAESTRO} Dataset},
  author={Curtis Hawthorne and Andriy Stasyuk and Adam Roberts and Ian Simon and Cheng-Zhi Anna Huang and Sander Dieleman and Erich Elsen and Jesse Engel and Douglas Eck},
  booktitle={International Conference on Learning Representations},
  year={2019},
  url={https://openreview.net/forum?id=r1lYRjC9F7},
}

@misc{kulkarni2026closerlookfailuremodes,
      title={A Closer Look at Failure Modes in Temporal Understanding of Large Audio-Language Models}, 
      author={Apoorva Kulkarni and Kaousheik Jayakumar and Sreyan Ghosh and Sarah Wiegreffe and Dinesh Manocha and Ramani Duraiswami},
      year={2026},
      eprint={2606.17417},
      archivePrefix={arXiv},
      primaryClass={cs.SD},
      url={https://arxiv.org/abs/2606.17417}, 
}

@article{Mesaros_2021,
   title={Sound Event Detection: A tutorial},
   volume={38},
   ISSN={1558-0792},
   url={http://dx.doi.org/10.1109/MSP.2021.3090678},
   DOI={10.1109/msp.2021.3090678},
   number={5},
   journal={IEEE Signal Processing Magazine},
   publisher={Institute of Electrical and Electronics Engineers (IEEE)},
   author={Mesaros, Annamaria and Heittola, Toni and Virtanen, Tuomas and Plumbley, Mark D.},
   year={2021},
   month=Sept, pages={67–83} }

@article{Politis_2021,
   title={Overview and Evaluation of Sound Event Localization and Detection in DCASE 2019},
   volume={29},
   ISSN={2329-9304},
   url={http://dx.doi.org/10.1109/TASLP.2020.3047233},
   DOI={10.1109/taslp.2020.3047233},
   journal={IEEE/ACM Transactions on Audio, Speech, and Language Processing},
   publisher={Institute of Electrical and Electronics Engineers (IEEE)},
   author={Politis, Archontis and Mesaros, Annamaria and Adavanne, Sharath and Heittola, Toni and Virtanen, Tuomas},
   year={2021},
   pages={684–698} }

@inproceedings{bredin23_interspeech,
  title     = {{pyannote.audio 2.1 speaker diarization pipeline: principle, benchmark, and recipe}},
  author    = {Hervé Bredin},
  year      = {2023},
  booktitle = {{Interspeech 2023}},
  pages     = {1983--1987},
  doi       = {10.21437/Interspeech.2023-105},
  issn      = {2958-1796},
}

@misc{wu2024largescalecontrastivelanguageaudiopretraining,
      title={Large-scale Contrastive Language-Audio Pretraining with Feature Fusion and Keyword-to-Caption Augmentation}, 
      author={Yusong Wu and Ke Chen and Tianyu Zhang and Yuchen Hui and Marianna Nezhurina and Taylor Berg-Kirkpatrick and Shlomo Dubnov},
      year={2024},
      eprint={2211.06687},
      archivePrefix={arXiv},
      primaryClass={cs.SD},
      url={https://arxiv.org/abs/2211.06687}, 
}

@misc{xu2023efficientsequencetransductionjointly,
      title={Efficient Sequence Transduction by Jointly Predicting Tokens and Durations}, 
      author={Hainan Xu and Fei Jia and Somshubra Majumdar and He Huang and Shinji Watanabe and Boris Ginsburg},
      year={2023},
      eprint={2304.06795},
      archivePrefix={arXiv},
      primaryClass={eess.AS},
      url={https://arxiv.org/abs/2304.06795}, 
}

@inproceedings{inproceedings,
author = {Mauch, Matthias and Dixon, Simon},
year = {2010},
month = {01},
pages = {135-140},
title = {Approximate Note Transcription for the Improved Identification of Difficult Chords.},
booktitle = {ISMIR 2010}
}

@misc{böck2016madmomnewpythonaudio,
      title={madmom: a new Python Audio and Music Signal Processing Library}, 
      author={Sebastian Böck and Filip Korzeniowski and Jan Schlüter and Florian Krebs and Gerhard Widmer},
      year={2016},
      eprint={1605.07008},
      archivePrefix={arXiv},
      primaryClass={cs.SD},
      url={https://arxiv.org/abs/1605.07008}, 
}

@misc{chen2022beatsaudiopretrainingacoustic,
      title={BEATs: Audio Pre-Training with Acoustic Tokenizers}, 
      author={Sanyuan Chen and Yu Wu and Chengyi Wang and Shujie Liu and Daniel Tompkins and Zhuo Chen and Furu Wei},
      year={2022},
      eprint={2212.09058},
      archivePrefix={arXiv},
      primaryClass={eess.AS},
      url={https://arxiv.org/abs/2212.09058}, 
}

@article{comanici2025gemini,
  title={Gemini 2.5: Pushing the frontier with advanced reasoning, multimodality, long context, and next generation agentic capabilities},
  author={Comanici, Gheorghe and Bieber, Eric and Schaekermann, Mike and Pasupat, Ice and Sachdeva, Noveen and Dhillon, Inderjit and Blistein, Marcel and Ram, Ori and Zhang, Dan and Rosen, Evan and others},
  journal={arXiv preprint arXiv:2507.06261},
  year={2025}
}

@misc{wang2024leveraginglanguagemodelcapabilities,
      title={Leveraging Language Model Capabilities for Sound Event Detection}, 
      author={Hualei Wang and Jianguo Mao and Zhifang Guo and Jiarui Wan and Hong Liu and Xiangdong Wang},
      year={2024},
      eprint={2308.11530},
      archivePrefix={arXiv},
      primaryClass={cs.SD},
      url={https://arxiv.org/abs/2308.11530}, 
}

@misc{yang2026multidomainaudioquestionanswering,
      title={Multi-Domain Audio Question Answering Benchmark Toward Acoustic Content Reasoning}, 
      author={Chao-Han Huck Yang and Sreyan Ghosh and Qing Wang and Jaeyeon Kim and Hengyi Hong and Sonal Kumar and Guirui Zhong and Zhifeng Kong and S Sakshi and Vaibhavi Lokegaonkar and Oriol Nieto and Ramani Duraiswami and Dinesh Manocha and Gunhee Kim and Jun Du and Rafael Valle and Bryan Catanzaro},
      year={2026},
      eprint={2505.07365},
      archivePrefix={arXiv},
      primaryClass={cs.SD},
      url={https://arxiv.org/abs/2505.07365}, 
}

@misc{bhattacharya2025benchmarkingconfidenceevaluationlalms,
      title={Benchmarking and Confidence Evaluation of LALMs For Temporal Reasoning}, 
      author={Debarpan Bhattacharya and Apoorva Kulkarni and Sriram Ganapathy},
      year={2025},
      eprint={2505.13115},
      archivePrefix={arXiv},
      primaryClass={cs.CL},
      url={https://arxiv.org/abs/2505.13115}, 
}

@misc{yin2026speakerlmendtoendversatilespeaker,
      title={SpeakerLM: End-to-End Versatile Speaker Diarization and Recognition with Multimodal Large Language Models}, 
      author={Han Yin and Yafeng Chen and Chong Deng and Luyao Cheng and Hui Wang and Chao-Hong Tan and Qian Chen and Wen Wang and Xiangang Li},
      year={2026},
      eprint={2508.06372},
      archivePrefix={arXiv},
      primaryClass={cs.SD},
      url={https://arxiv.org/abs/2508.06372}, 
}

@phdthesis{raffel2016learning,
  title     = {Learning-Based Methods for Comparing Sequences, with Applications to Audio-to-MIDI Alignment and Matching},
  author    = {Raffel, Colin},
  school    = {Columbia University},
  year      = {2016}
}

@inproceedings{radford2023robust,
  title={Robust speech recognition via large-scale weak supervision},
  author={Radford, Alec and Kim, Jong Wook and Xu, Tao and Brockman, Greg and McLeavey, Christine and Sutskever, Ilya},
  booktitle={International conference on machine learning},
  pages={28492--28518},
  year={2023},
  organization={PMLR}
}

@misc{shao2024deepseekmath,
      title={DeepSeekMath: Pushing the Limits of Mathematical Reasoning in Open Language Models}, 
      author={Zhihong Shao and Peiyi Wang and Qihao Zhu and Runxin Xu and Junxiao Song and Xiao Bi and Haowei Zhang and Mingchuan Zhang and Y. K. Li and Y. Wu and Daya Guo},
      year={2024},
      eprint={2402.03300},
      archivePrefix={arXiv},
      primaryClass={cs.CL},
      url={https://arxiv.org/abs/2402.03300}, 
}

@misc{wu2025audiothinkerguidingaudiolanguage,
      title={Audio-Thinker: Guiding Audio Language Model When and How to Think via Reinforcement Learning}, 
      author={Shu Wu and Chenxing Li and Wenfu Wang and Hao Zhang and Hualei Wang and Meng Yu and Dong Yu},
      year={2025},
      eprint={2508.08039},
      archivePrefix={arXiv},
      primaryClass={cs.SD},
      url={https://arxiv.org/abs/2508.08039}, 
}

@misc{li2025reinforcementlearningoutperformssupervised,
      title={Reinforcement Learning Outperforms Supervised Fine-Tuning: A Case Study on Audio Question Answering}, 
      author={Gang Li and Jizhong Liu and Heinrich Dinkel and Yadong Niu and Junbo Zhang and Jian Luan},
      year={2025},
      eprint={2503.11197},
      archivePrefix={arXiv},
      primaryClass={cs.SD},
      url={https://arxiv.org/abs/2503.11197}, 
}

@INPROCEEDINGS{7178964,
  author={Panayotov, Vassil and Chen, Guoguo and Povey, Daniel and Khudanpur, Sanjeev},
  booktitle={2015 IEEE International Conference on Acoustics, Speech and Signal Processing (ICASSP)}, 
  title={Librispeech: An ASR corpus based on public domain audio books}, 
  year={2015},
  volume={},
  number={},
  pages={5206-5210},
  doi={10.1109/ICASSP.2015.7178964}}

@inproceedings{piczak2015dataset,
  title = {{ESC}: {Dataset} for {Environmental Sound Classification}},
  author = {Piczak, Karol J.},
  booktitle = {Proceedings of the 23rd {Annual ACM Conference} on {Multimedia}},
  date = {2015-10-13},
  url = {http://dl.acm.org/citation.cfm?doid=2733373.2806390},
  doi = {10.1145/2733373.2806390},
  location = {{Brisbane, Australia}},
  isbn = {978-1-4503-3459-4},
  publisher = {{ACM Press}},
  pages = {1015--1018},
  year={2015}
}

@misc{tang2024salmonngenerichearingabilities,
      title={SALMONN: Towards Generic Hearing Abilities for Large Language Models}, 
      author={Changli Tang and Wenyi Yu and Guangzhi Sun and Xianzhao Chen and Tian Tan and Wei Li and Lu Lu and Zejun Ma and Chao Zhang},
      year={2024},
      eprint={2310.13289},
      archivePrefix={arXiv},
      primaryClass={cs.SD},
      url={https://arxiv.org/abs/2310.13289}, 
}

@misc{chu2024qwen2audiotechnicalreport,
      title={Qwen2-Audio Technical Report}, 
      author={Yunfei Chu and Jin Xu and Qian Yang and Haojie Wei and Xipin Wei and Zhifang Guo and Yichong Leng and Yuanjun Lv and Jinzheng He and Junyang Lin and Chang Zhou and Jingren Zhou},
      year={2024},
      eprint={2407.10759},
      archivePrefix={arXiv},
      primaryClass={eess.AS},
      url={https://arxiv.org/abs/2407.10759}, 
}

\appendix
\section{Dataset and Preprocessing Details}
\label{app:datasets}

\paragraph{Multi-Speaker ASR and Diarization.} AMI, ICSI, and Switchboard feature
spontaneous, overlapping, naturalistic speech. Recordings are segmented into chunks of
up to 60 seconds while preserving multi-speaker structure within each chunk.

\paragraph{Audio Temporal Grounding.} AudioSet Strong provides categorical event labels
rather than natural-language queries. We use
GPT-4o~\citep{openai2024gpt4technicalreport} to generate diverse paraphrased queries
for each event class (e.g., for \texttt{dog\_bark}: ``When does a dog start
barking?''), with the strong temporal labels serving as ground-truth intervals.

\paragraph{Dense Audio Captioning.} TACOS contains 12,358 FreeSound recordings
(15--30 seconds each) annotated with 47,748 timestamped captions.

\paragraph{Timestamped Music Captioning.} Slakh2100 contains 145 hours of audio split
into 30-second segments. Chord labels are extracted by quantizing simultaneously
sounding notes into standard chord symbols (e.g., C major, G7); instrument
entries/exits, tempo, dynamics, and note statistics are read directly from the MIDI
metadata.

\paragraph{Synthetic speech compositions.} We synthesize multi-speaker audio by
randomly selecting 1-4 speakers from LibriSpeech and concatenating their utterances
onto a shared 60-second canvas with small gaps between segments. Speaker slots are
balanced across 1-4 speakers in equal proportions. Segment start and end times are
chosen to avoid over-representing any particular timestamp value, and the final audio
is peak-normalized.

\paragraph{Synthetic audio event compositions.} We synthesize audio event recordings
by randomly placing ESC-50 sound clips onto a 60-second canvas, with each composition
containing 1-6 overlapping or non-overlapping events drawn from different sound
categories. Separate pools are generated for DAC and AG to decouple their category
distributions. For AG, we ensure at least 50\% of compositions contain repeated
occurrences of the same sound category, so that the model must learn to localize
multiple spans of the same event. A random 10\% of AG compositions ground the full
clip rather than a specific event. As with speech, timestamp values are balanced
uniformly across all compositions.
\section{Evaluation Metric Details}
\label{app:eval_details}

This appendix gives the full computational definitions of the evaluation metrics. All interval matching uses the greedy IoU-based matching described below unless stated otherwise.

\paragraph{Word Error Rate (WER).} For multi-speaker ASR, we first greedily match predicted segments to ground-truth segments by temporal IoU. For each matched pair we compute the word-level Levenshtein distance between the normalized predicted and reference transcripts, and accumulate it into a total edit count. Unmatched ground-truth segments contribute all of their reference words as deletions, and unmatched predicted segments contribute all of their words as insertions. The final WER is the total edit count divided by the total reference word count across all segments.

\paragraph{Diarization Error Rate (DER).} We use the standard DER as implemented in \texttt{pyannote.metrics}~\citep{bredin17_interspeech}, with a collar of $0.25$~s. Speaker labels are optimally permuted between hypothesis and reference before scoring, so that a hypothesis with consistent but arbitrarily named speakers incurs no penalty from the labeling alone. DER aggregates missed speech, false alarms, and speaker confusion as a fraction of total reference speech duration.

\paragraph{Symmetric mIoU.} For a set of reference intervals and a set of predicted intervals, we define the recall-side mean-best-IoU as the average over reference intervals of the maximum IoU with any predicted interval, and the precision-side mean-best-IoU symmetrically as the average over predicted intervals of the maximum IoU with any reference interval. The reported symmetric mIoU is the harmonic mean of these two quantities. The harmonic mean prevents over-prediction from inflating the score: emitting many spurious intervals raises recall-side IoU but lowers precision-side IoU, and the harmonic mean is dominated by the smaller of the two.

\paragraph{F1 at IoU threshold.} For dense audio captioning and audio grounding, a predicted interval is a true positive if its IoU with a greedily-matched ground-truth interval is at least $0.5$; precision, recall, and F1 are computed from the resulting true-positive, false-positive, and false-negative counts. For diarization, the speaker-labeled F1@IoU0.5 additionally requires that the predicted and ground-truth segments carry the same speaker label, so a temporally correct segment with the wrong speaker is not counted as a true positive.

\paragraph{Boundary MAE.} For tasks with continuous time boundaries (ASR, grounding, dense captioning, music), we report the mean absolute error of segment onsets and offsets in seconds. The MAE is computed only over greedy-matched pairs with $\mathrm{IoU} \geq 0.3$, so that boundary error is measured only where the prediction and reference refer to the same underlying event.

\paragraph{METEOR.} For dense audio captioning, caption quality is measured with METEOR computed over greedy-matched predicted and ground-truth segment pairs and averaged across matched pairs. Matching uses the same IoU-based greedy procedure as the other interval metrics.

\paragraph{Music-specific metrics.} For chords, we report two frame-level accuracies sampled at $0.5$~s resolution: \emph{root accuracy} (fraction of sampled frames with the correct chord root) and \emph{quality accuracy} (fraction with the correct chord quality label). We additionally report \emph{chord F1@IoU0.5}, which greedily matches predicted and ground-truth chord segments requiring both $\mathrm{IoU} \geq 0.5$ and an exact chord-label match, and \emph{chord mIoU}, the symmetric mIoU over all chord segments. For instruments, we report \emph{instrument-name F1} (token-level F1 between the predicted and ground-truth instrument-name sets) and \emph{instrument boundary MAE} (onset/offset error in seconds over matched instruments). \emph{Tempo accuracy} is the fraction of examples whose predicted BPM falls within $4\%$ of the ground truth.
\section{Per-Component SFT Ablation}
\label{app:component_ablation}

\begin{table*}[t]
\centering
\small
\caption{SFT component ablation on a controlled 2{,}000-example subset, reported on three representative tasks. Loss: distance-aware Gaussian timestamp loss. Proj: time-aware multi-modal projector. DER, eF1, F1, mIoU, and MET (METEOR) are reported as percentages; MAE in seconds. $\downarrow$: lower is better; $\uparrow$: higher is better. Best result per column in \textbf{bold}.}
\label{tab:component_ablation}
\setlength{\tabcolsep}{5pt}
\renewcommand{\arraystretch}{1.2}
\begin{tabular}{@{}l cc ccc ccc@{}}
\toprule
\multirow{2}{*}{\textbf{Variant}}
  & \multicolumn{2}{c}{\textbf{Diarization}}
  & \multicolumn{3}{c}{\textbf{Dense Captioning}}
  & \multicolumn{3}{c}{\textbf{Audio Grounding}} \\
\cmidrule(lr){2-3} \cmidrule(lr){4-6} \cmidrule(lr){7-9}
  & DER$\downarrow$ & mIoU$\uparrow$
  & eF1$\uparrow$ & mIoU$\uparrow$ & MET$\uparrow$
  & F1$\uparrow$ & MAE$\downarrow$ & mIoU$\uparrow$ \\
\midrule
Naive          & 98.4           & 23.8          & 35.3          & 43.3          & 20.5          & 33.9          & 2.05          & 36.1          \\
+ Loss         & 86.2           & \textbf{37.5} & \textbf{49.9} & \textbf{60.2} & 21.5          & 38.0          & 1.73          & 41.9          \\
+ Projector    & 104.2          & 22.2          & 34.2          & 42.8          & 20.5          & 34.0          & 2.02          & 36.4          \\
+ Both (Ours)  & \textbf{79.3}  & \textbf{37.5} & 49.7          & 60.1          & \textbf{21.9} & \textbf{38.7} & \textbf{1.69} & \textbf{42.5} \\
\bottomrule
\end{tabular}
\end{table*}

To isolate the contribution of each SFT innovation, we train four variants on a controlled subset of 2{,}000 real-world examples: (i) \emph{Naive}, with neither the distance-aware loss nor the time-aware projector; (ii) \emph{+ Loss}, adding only the distance-aware Gaussian loss; (iii) \emph{+ Projector}, adding only the time-aware multi-modal projector; and (iv) \emph{+ Both}, the full SFT recipe with both components. Due to compute constraints, we report three representative tasks spanning speech and audio (diarization, dense audio captioning, and audio temporal grounding) in Table~\ref{tab:component_ablation}.

An interesting pattern emerges. The distance-aware loss, in isolation, provides large improvements across all three tasks, including a 14.6-point gain in dense captioning event F1 and a corresponding gain in diarization speaker-labeled F1 and grounding F1. In contrast, the time-aware projector in isolation is not useful, suggesting that wall-clock positional information requires a corresponding learning signal to be effective. However, when paired with the loss, the projector contributes a clear additional benefit on the two tasks with the densest temporal structure: diarization (DER improves from 86.2\% to 79.3\%) and audio grounding (F1 from 38.0\% to 38.7\%, MAE from 1.73~s to 1.69~s, mIoU from 41.9\% to 42.5\%).
\section{Baseline Prompts and Format failure analysis}
\label{app:baseline_failure_analysis}

\paragraph{Tolerant output parsing.} The zero-shot LALMs use different native timestamp conventions. AF3 and AF-Next are prompted using the \texttt{<t>start-end</t> text} format, whereas Qwen3-Omni uses timestamp tokens of the form \texttt{ <|start|> text <|end|>}. We therefore use a separate, deliberately tolerant parser for each model. The parsers accept common model-specific variants, including bracketed intervals, paired timestamp tokens, silence markers, and speaker-label variants. Outputs from which no timestamped interval can be recovered are treated as format failures and scored as maximally incorrect rather than being discarded. For each zero-shot baseline, we evaluated three closely related prompt
scaffolds and report the best-performing configuration; Exact baseline prompts are provided in Table \ref{tab:baseline_prompts}. 

\paragraph{Format validation.} To determine whether poor zero-shot performance is primarily caused by failure to follow the requested output syntax, we recompute ASR and diarization metrics using only examples with parseable baseline outputs. For comparison, TEMPO is evaluated on exactly the same valid-example subset for each baseline. \begin{table}[t] \centering \small \caption{ASR format-validation analysis. ``Valid'' denotes evaluation after excluding outputs from which no timestamped interval can be recovered. TEMPO is evaluated on the identical subset. WER is reported in \%.} \label{tab:asr_format_validation} \setlength{\tabcolsep}{3.5pt} \begin{tabular}{lrrrr} \toprule \textbf{Model} & \textbf{Invalid} & \textbf{WER} & \textbf{WER} & \textbf{TEMPO} \\ & \textbf{(\%)} & \textbf{all} & \textbf{valid} & \textbf{valid} \\ \midrule Audio Flamingo 3 & 5.1 & 209.7 & 206.7 & 44.7 \\ Audio Flamingo Next & 5.8 & 167.1 & 162.2 & 44.1 \\ Qwen3-Omni & 18.9 & 69.7 & 62.7 & 45.5 \\ \bottomrule \end{tabular} \end{table} \begin{table}[t] \centering \small \caption{Diarization format-validation analysis. DER is reported in \%. TEMPO is evaluated on the same valid-example subset as each baseline.} \label{tab:diar_format_validation} \setlength{\tabcolsep}{3.5pt} \begin{tabular}{lrrrr} \toprule \textbf{Model} & \textbf{Invalid} & \textbf{DER} & \textbf{DER} & \textbf{TEMPO} \\ & \textbf{(\%)} & \textbf{all} & \textbf{valid} & \textbf{valid} \\ \midrule Audio Flamingo 3 & 21.7 & 103.2 & 101.2 & 27.2 \\ Audio Flamingo Next & 23.6 & 106.9 & 103.6 & 31.8 \\ Qwen3-Omni & 0.0 & 44.2 & 44.2 & 28.0 \\ \bottomrule \end{tabular} \end{table} Restricting evaluation to parseable outputs changes the baseline scores only modestly. For example, AF3's ASR WER decreases from 209.7\% to 206.7\%, while its diarization DER remains above 100\% (103.2\% to 101.2\%) even when all format failures are excluded. Similarly, Qwen3-Omni has no diarization format failures, yet its DER remains 44.2\%, compared with 28.0\% for TEMPO on the same examples. Thus, the large baseline errors primarily reflect incorrect content and temporal predictions rather than strict output parsing. 

\paragraph{Failure analysis.} TEMPO's remaining errors are dominated by temporal granularity and segment-boundary mistakes rather than degenerate generations. In dense audio captioning, the 10.2\% of examples with severe temporal failure ($\mathrm{mIoU}<0.1$) still obtain a METEOR score of 0.28, compared with a dataset-wide average of 0.23, indicating that the event descriptions are often semantically correct even when their temporal boundaries are inaccurate. For multi-speaker ASR, only 7.1\% of examples have WER above 100\%; these examples also exhibit substantially lower segment overlap (mean IoU 0.34 versus 0.58 overall), suggesting that the dominant failure mode is merging adjacent speaker turns or misplacing their boundaries. The zero-shot baselines exhibit qualitatively different failures. AF3 frequently generates plausible but unsupported dialogue or caption-like content instead of faithful transcription, AF-Next can enter repetitive generation loops, and Qwen3-Omni sometimes collapses an entire clip into a single untimestamped segment. These behaviors explain why WER and DER can exceed 100\% even when the outputs are otherwise parseable. 

\begin{table*}[t]
\centering
\scriptsize
\caption{
Exact task-specific prompt suffixes used for the zero-shot LALM
baselines. For each example, the corresponding evaluation question is
followed by the model-specific suffix shown below. No system prompt is
used for these models.
}
\label{tab:baseline_prompts}
\setlength{\tabcolsep}{4pt}
\renewcommand{\arraystretch}{1.15}

\begin{tabularx}{\textwidth}{
>{\raggedright\arraybackslash}p{0.09\textwidth}
>{\raggedright\arraybackslash}X
>{\raggedright\arraybackslash}X
>{\raggedright\arraybackslash}X}
\toprule
\textbf{Task}
& \textbf{Audio Flamingo 3}
& \textbf{Audio Flamingo Next}
& \textbf{Qwen3-Omni} \\
\midrule

\textbf{ASR}
&
\texttt{
Transcribe the speech with timestamps. Break it into short segments
(a few words each). Use <t>start-end</t> format for each segment,
one per line.
Example:
<t>0.0-2.5</t> Hello world
<t>3.1-4.8</t> How are you
<t>5.2-7.0</t> I am fine thank you
Provide accurate start and end times in seconds.
}
&
\texttt{
Transcribe the input speech. If multiple speakers are present, provide
diarized transcripts with speaker labels. Include timestamps for each
utterance using <t>start-end</t> format.
Example:
<t>0.0-2.5</t> [Speaker 1] Hello world
<t>3.1-4.8</t> [Speaker 2] How are you
<t>5.2-7.0</t> [Speaker 1] I am fine thank you
Break into multiple short segments with accurate timestamps.
}
&
\texttt{
Respond using ONLY the following format. Split the transcription into
SHORT segments (a few words each) with precise start AND end timestamps
in seconds. Each segment MUST have both <|start|> and <|end|>
timestamps. Output one segment per line:
[speech:asr] <|start\_seconds|> transcribed text <|end\_seconds|>
Example:
[speech:asr] <|0.0|> Hello world <|2.5|>
[speech:asr] <|3.1|> How are you <|4.8|>
[speech:asr] <|5.2|> I am fine thank you <|7.0|>
Do NOT put all text into a single segment. Break it into multiple short
segments with accurate timestamps.
}
\\
\midrule

\textbf{Diarization}
&
\texttt{
Identify each speaker and when they speak. Use <t>start-end</t> format
with speaker labels, one turn per line.
Example:
<t>0.0-3.5</t> [Speaker 1] okay hi
<t>3.8-6.2</t> [Speaker 2] hello how are you
<t>6.5-9.0</t> [Speaker 1] i'm good thanks
}
&
\texttt{
Transcribe the input audio. If multiple speakers are present, provide
diarized transcripts with speaker labels. Include timestamps for each
speaker turn using <t>start-end</t> format.
Example:
<t>0.0-3.5</t> [Speaker 1] okay hi
<t>3.8-6.2</t> [Speaker 2] hello how are you
<t>6.5-9.0</t> [Speaker 1] i'm good thanks
}
&
\texttt{
Respond using ONLY the following format, one segment per line:
[speech:diar] <|start\_seconds|> Speaker N <|end\_seconds|>
where N is the speaker number (1, 2, 3, ...).
Example:
[speech:diar] <|0.0|> Speaker 1 <|3.5|>
[speech:diar] <|3.8|> Speaker 2 <|6.2|>
}
\\
\midrule

\textbf{Dense Captioning}
&
\texttt{
Write a time-segmented caption of the audio, capturing sound, speech,
and music details. Use <t>start-end</t> format for each event, one per
line.
Example:
<t>0-3</t> A dog barks loudly in the background.
<t>2-8</t> Wind blows steadily outdoors.
<t>5-10</t> A car engine revs and drives away.
}
&
\texttt{
Write a time segmented caption of the input audio capturing sound,
speech and music details. Use <t>start-end</t> format for each segment.
Example:
<t>0-3</t> A dog barks loudly in the background.
<t>2-8</t> Wind blows steadily outdoors.
<t>5-10</t> A car engine revs and drives away.
}
&
\texttt{
Respond using ONLY the following format, one event per line:
[audio:caption] <|start\_seconds|> description of sound event
<|end\_seconds|>
Example:
[audio:caption] <|0.0|> A dog barks loudly. <|3.5|>
[audio:caption] <|2.1|> Wind blows in the background. <|8.0|>
}
\\
\midrule

\textbf{Audio Grounding}
&
\texttt{
Provide the exact time intervals where the queried sound event occurs.
Use <t>start-end</t> format, one interval per line.
Example:
<t>0.9-1.4</t>
<t>3.7-4.3</t>
<t>6.5-7.5</t>
}
&
\texttt{
Provide the exact time intervals where the queried sound event occurs
in the audio. Use <t>start-end</t> format, one interval per line.
Example:
<t>0.9-1.4</t>
<t>3.7-4.3</t>
<t>6.5-7.5</t>
}
&
\texttt{
Respond using ONLY the following format, one interval per line:
[audio:ground] <|start\_seconds|> to <|end\_seconds|>
Example:
[audio:ground] <|0.9|> to <|1.4|>
[audio:ground] <|3.7|> to <|4.3|>
}
\\
\midrule

\textbf{Music Captioning}
&
\texttt{
Describe the music in detail: instruments, tempo (BPM), key, chord
progressions, dynamics, and note statistics. Include timestamps using
<t>start-end</t> format. Mention style, arrangement, production, and
emotions conveyed.
}
&
\texttt{
Summarize the track with precision: mention its musical style, BPM,
key, arrangement, production choices, and the emotions or story it
conveys. Include timestamps using <t>start-end</t> format. Also
describe instruments, tempo changes, chord progressions, dynamics, and
note statistics (density and range).
}
&
\texttt{
Respond using ONLY the following format on a single line:
[instrument] InstrumentName enters at <|start|> exits at <|end|>
[tempo] BPM at <|time|>
[chord] ChordName from <|start|> to <|end|>
[dynamics] description
[note] density: X range: NoteA to NoteB
Example:
[instrument] Piano enters at <|0.00|> exits at <|10.00|>
[tempo] 120.0 BPM at <|0.00|>
[chord] C from <|0.00|> to <|2.50|>
[chord] G from <|2.50|> to <|5.00|>
[dynamics] moderate and steady
[note] density: 4.5 range: C3 to C5
}
\\

\bottomrule
\end{tabularx}
\end{table*}
\section{Qualitative Diagnostic Analysis}
\label{app:qualitative_analysis}

\begin{figure*}[t]
    \centering
    \includegraphics[width=0.95\textwidth]{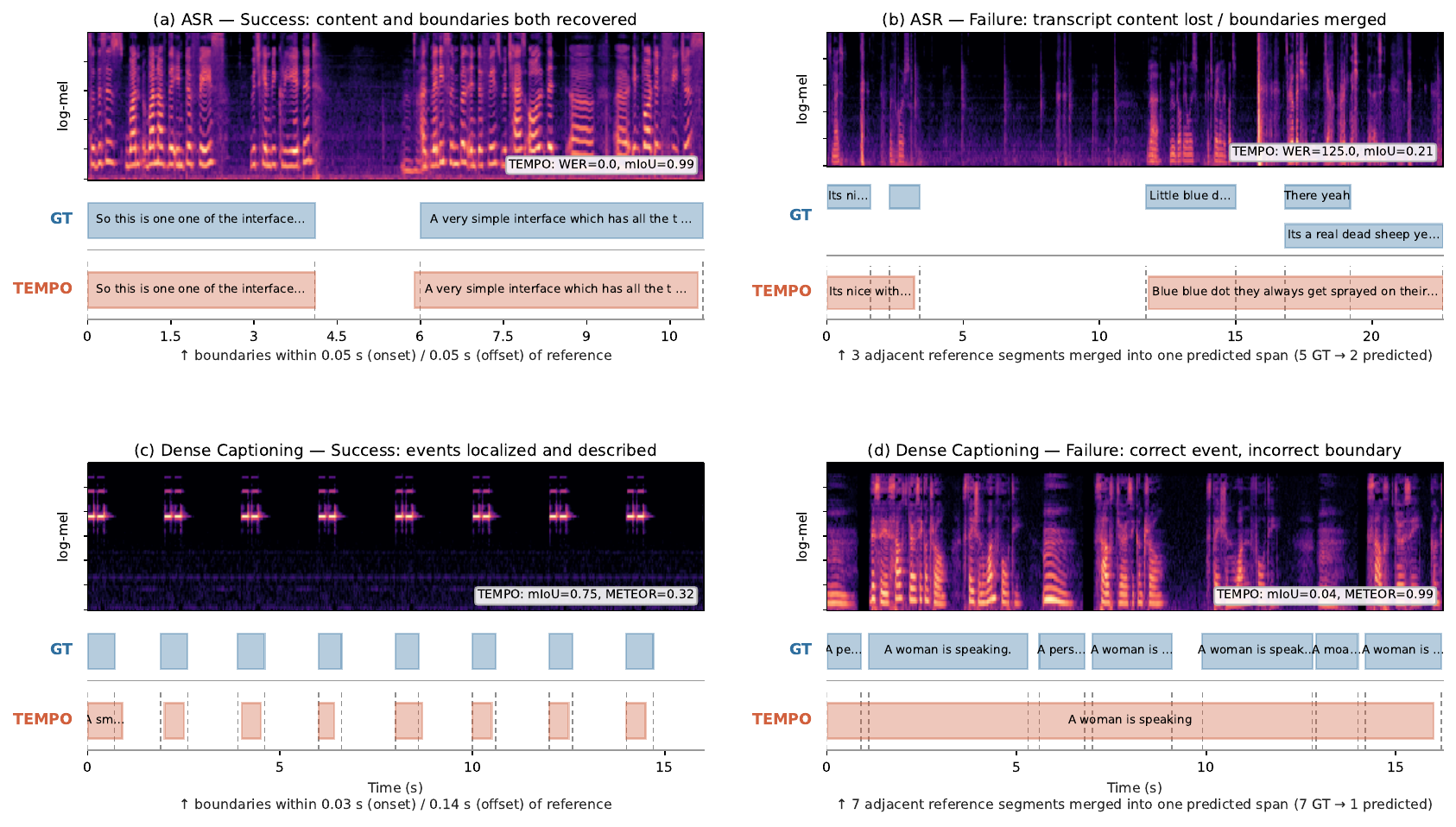}
    \caption{\small
    Qualitative diagnostic examples for ASR and dense audio captioning.
    Each panel shows the log-mel spectrogram together with ground-truth
    intervals and TEMPO predictions. \textbf{(a)} ASR success with
    accurate transcript content and temporal boundaries. \textbf{(b)}
    ASR failure in which adjacent reference utterances are merged.
    \textbf{(c)} Dense-captioning success with repeated events
    individually localized. \textbf{(d)} Dense-captioning failure in
    which seven reference events collapse into a single prediction
    despite high semantic agreement.}
    \label{fig:qualitative_speech_sound}
    \vspace{-5mm}
\end{figure*}

Aggregate metrics do not reveal whether poor temporal overlap arises
from incorrect content, collapsed temporal structure, or boundary
misalignment. We therefore inspect representative held-out examples
and visualize the input log-mel spectrogram together with the
ground-truth and TEMPO prediction intervals. The examples are selected
using per-example task metrics, and all predictions are taken directly
from the final GRPO checkpoint without re-inference or manual editing.
Figure~\ref{fig:qualitative_speech_sound} presents representative
success and failure cases for ASR and dense audio captioning.

In Figure~\ref{fig:qualitative_speech_sound}\textbf{(a)}, TEMPO
accurately recovers both the ASR transcript and its temporal structure,
achieving 0 WER and 0.99 mIoU, with onset and offset errors below
0.05\,s. In \textbf{(b)}, five reference ASR segments collapse into
two predictions, including three adjacent utterances merged into a
single interval; consequently, WER rises to 125\% and mIoU falls to
0.21. In \textbf{(c)}, TEMPO successfully preserves the temporal
granularity of repeated sound events in dense audio captioning,
achieving 0.75 mIoU. In contrast, \textbf{(d)} shows a granularity
collapse in which seven reference events are merged into a single
predicted interval. Despite this severe temporal error, the predicted
event description remains semantically correct (METEOR = 0.99), while
mIoU drops to 0.04. These examples illustrate that TEMPO's residual
errors can arise from segment merging and boundary misalignment even
when the underlying semantic content is correctly identified.

\section{Potential Risks}

We do not foresee significant risks specific to this work beyond those already inherent to existing large audio-language models and speech recognition systems. \textsc{Tempo} performs temporal localization of events in audio and does not generate speech, identify individual speakers by name, or enable capabilities not already present in publicly available audio understanding systems. All training and evaluation data are drawn from established, publicly released research corpora.

\section{License for Artifacts}
\label{sec:licenses}

We use a number of publicly released scientific artifacts in this work. Below, we summarize the licenses under which they are distributed and confirm that our use is consistent with their intended terms.

\paragraph{Datasets.} AMI and ICSI are released under the CC BY 4.0 license for research use. Switchboard is distributed by the Linguistic Data Consortium (LDC) under a research license. AudioSet and AudioSet Strong are released under CC BY 4.0. TACOS is released under CC BY 4.0. Slakh2100 is released under CC BY 4.0. LibriSpeech is released under CC BY 4.0. ESC-50 is released under the CC BY-NC 3.0 license, which permits non-commercial research use.

\paragraph{Models.} Audio Flamingo 3, our base model, is released under a research-only license by NVIDIA. Baselines we evaluate against (Audio Flamingo Next, Qwen3-Omni, TimeAudio) are accessed through their official releases under their respective research licenses.

\paragraph{Compatibility with intended use.} All artifacts are used for non-commercial academic research, which is consistent with the intended use specified by each provider.

\section{Computational Resources}
\label{sec:compute}

\textsc{Tempo} is built on Audio Flamingo 3, which consists of a frozen Whisper-large audio encoder (approximately 0.6B parameters) and a Qwen2-7B language model (approximately 7B parameters), connected by a two-layer MLP projector. The audio encoder remains frozen throughout training.
All experiments are run on NVIDIA A6000 GPUs. SFT runs on 8 GPUs on a single node, with Stage 1 covering 51{,}512 synthetic examples over two epochs and Stage 2 covering 32{,}726 real examples over two additional epochs. GRPO training (1{,}000 steps with 8 completions per prompt) runs on 8 GPUs distributed across 2 nodes (4 GPUs per node) using DDP via HuggingFace Accelerate. Evaluation across all five tasks (10{,}513 question-answer pairs) is run on a single A6000.

\section{Use of AI Assistants}
\label{app:ai_use}
We utilized AI assistants to help clarify explanations, suggest concise phrasing, and organize text for readability. These tools were used exclusively for linguistic support and were not used to generate scientific results or formulate claims. 

\section*{Acknowledgement} 
The research is partially supported by Adobe, Amazon, NVIDIA, and Sesame

\end{document}